\documentclass[10pt, emptycopyrightspace]{ewsn-proc}

\usepackage{tikz}
\usepackage{amsmath}

\usepackage{orcidlink}
\usepackage[tight,footnotesize]{subfigure}
\usepackage{float}
\usepackage{graphicx}
\usepackage{xcolor,colortbl}
\usepackage{tcolorbox}
\usepackage[english]{babel}
\usepackage[utf8]{inputenc}
\usepackage{algorithm}
\usepackage{balance} 
\usepackage{bm}
\usepackage[nolist,nohyperlinks]{acronym}
\usepackage[noend]{algpseudocode}
\usepackage[inline]{enumitem}
\usepackage{url}
\usepackage{authblk}
\def\checkmark{\tikz\fill[scale=0.4](0,.35) -- (.25,0) -- (1,.7) -- (.25,.15) -- cycle;} 
\let\oldReturn\Return
\renewcommand{\Return}{\State\oldReturn}

\definecolor{green}{rgb}{0.1,0.1,0.1}

\graphicspath{ {./} }
\newcommand{\ourmethod}{\textit{FedPut}}

\newcommand{\argmin}[1]{\underset{#1}{\operatorname{arg}\,\operatorname{min}}\;}

\tcbuselibrary{theorems}
\newtcbtheorem
  []
  {takeaway}
  {Takeaway}
  {%
    colback=green!5,
    colframe=green!35!black,
    fonttitle=\bfseries,
  }
  {tay}

\author{
 \alignauthor Argha Sen$^{\dag}$, Ayan Zunaid$^{\dag *}$, Soumyajit Chatterjee$^{\dag}$\thanks{\text{ }Author was a student at IIT Kharagpur when this work was completed.}\text{ }, Basabdatta Palit$^{\mathsection}$, Sandip Chakraborty$^{\dag}$ \\
\affaddr{$^{\dag}$IIT Kharagpur, India, $^{\mathsection}$NIT Rourkela, India} \\
\email{\{arghasen10, ayanzunaid10, sjituit\}@gmail.com, \text{palitb@nitrkl.ac.in}, sandipc@cse.iitkgp.ac.in
}
}

\makeatletter
\def\@copyrightspace{\relax}
\makeatother

\title{Revisiting Cellular Throughput Prediction over the Edge: Collaborative Multi-device, Multi-network \textit{in-situ} Learning}
\begin{document}



\maketitle


\begin{abstract}
Pervasive applications over large-scale, distributed embedded devices and the Internet of Things (IoT) demand precise coordination with the network; for example, several such applications, like collaborative video streaming and live analysis, augmented reality, etc., need continuous monitoring of network throughput and adapt the application behavior accordingly. Although the idea of network throughput prediction is not new and quite dated, in this paper, we show that the existing approaches fail to correctly infer the throughput when the network operator or the device change, and thus, not generic enough for Internet-scale applications. We propose \ourmethod, a novel approach that allows collaborative training across different client hardware by capturing throughput variations based on devices' sensitivity towards the corresponding network configurations. Rigorous evaluations show that \ourmethod{} outperforms various standard baseline algorithms with more than $80\%$ R2-score over different datasets. We also analyze the performance of \ourmethod{} over a network-aware streaming media application and demonstrate its efficacy for various application scenarios. 

\end{abstract}

\begin{acronym}
	\acro{2G}{2$^\text{nd}$ Generation}
	\acro{3G}{3$^\text{rd}$ Generation}
	\acro{4G}{4$^\text{th}$ Generation}
	\acro{5G}{5$^\text{th}$ Generation}
	\acro{A3C}{Actor-Critic}
	\acro{ABR}{adaptive bitrate}
	\acro{BS}{Base Station}
	\acro{CDN}{Content Distribution Network}
	\acro{DASH}{Dynamic Adaptive Streaming over HTTP}
	\acro{DL}{deep learning}
	\acro{DRX}{Discontinuous Reception}
	\acro{EDGE}{Enhanced Data Rates for \ac{GSM} Evolution.}
			\acro{gNB}{general NodeB}
	\acro{eNB}{evolved NodeB}
	\acro{GSM}{Global System for Mobile}
	\acro{FL}{Federated Learning}
	\acro{TL}{Transfer Learning}
	\acro{HD}{High Definition}
	\acro{HSPA}{High Speed Packet Access}
	\acro{LSTM}{Long Short Term Memory}
	\acro{LTE}{Long Term Evolution}
	\acro{ML}{machine learning}
	\acro{MTL}{Multi-Task Learning}
	\acro{MCS}{Modulation and Coding Scheme}
	\acro{NSA}{Non-Standalone}
	\acro{HVPM}{High voltage Power Monitor}
	\acro{QoS}{Quality of Service}
	\acro{QoE}{Quality of Experience}
	\acro{RF}{Random Forest}
	\acro{RFL}{Random Forest}
	\acro{RL}{Reinforcement Learning}
	\acro{RRC}{Radio Resource Control}
	\acro{RSSI}{Received Signal Strength Indicator}
	\acro{RSRP}{Reference Signal Received Power}
	\acro{RSRQ}{Reference Signal Received Quality}
	\acro{SINR}{signal-to-interference-plus-noise-ratio}
	\acro{SNR}{signal-to-noise-ratio}
	\acro{UE}{User Equipment}
	\acro{UHD}{Ultra HD}
	\acro{VoLTE}{Voice over LTE}
	\acro{RNN}{Recurrent Neural Network}
	\acro{WiFi}{Wireless Fidelity}
	\acro{ARIMA}{Auto Regressive Integrated Moving Average}
	\acro{ML}{Machine Learning}
	\acro{NR}{New Radio}
	\acro{Non-IID} {Independent and Identically Distributed}
	\acro{TP}{Throughput Prediction}
	\acro{CQI}{Channel Quality Indicator}
	\acro{CTFL}{Cross-Technology Federated Learning}
        \acro{NAA}{Network Aware Application}
        \acro{SVR}{Support Vector Regression}
        \acro{MLP}{Multilayer Perceptrons}
\end{acronym}

\section{Introduction}
\label{sec_intro}
Rapid deployment of 5G networks has multiplied the demand for various large-scale, multi-connectivity, high bandwidth applications, such as live broadcast from thousands of cameras, collaborative high-definition video streaming, remote and visual inspections, facility management through augmented reality, connected vehicles, etc. These applications are required to deliver precise Quality of Experience (QoE) to the end-users over thousands of interconnected devices, for which the applications tune themselves to the condition of the underlying network. An example use case is Adaptive Bitrate (ABR) collaborative video processing and streaming over mobile edges~\cite{tran2018adaptive,tuysuz2020qoe,farahani2022ararat}, which tunes the video playback quality over the edge devices according to the network performance counters in order to meet the QoE requirements of the end-users. To enable such effective use of network-related information by different applications, several recent works have focused on designing APIs for \textit{Network-aware Application} (NAA), such as \textit{Mobile and Wireless Information Exposure} (MoWIE)~\cite{Zhang2020} and \textit{Network Exposure Functions} (NEF) for 5G networks. However, for NAA(s) to decide the state, a primary requirement is to estimate the perceived network throughput for the corresponding applications accurately. Notably, there is a plethora of studies conducted on throughput prediction for cellular networks~\cite{Issa2019, Bentaleb2019, Gao2021Hermes, Raca2019, Raca2018_2, Raca2020_1, Raca2020, raca2017back, Yue2018, yin2021ant}, albeit most of these rely on time series models~\cite{Issa2019}, linear regression, trained random forests and support vector regression~\cite{Raca2019,Raca2020,Yue2018}, as well as Deep Neural Networks (DNNs)~\cite{Wei2019,Narayanan2020, Raca2020_1} all of which are \textbf{trained in a supervised manner}. Subsequently, to accommodate such a supervised regime, these models are often trained using \textbf{datasets acquired in} \textbf{restricted settings} with different network parameters such as the signal quality, cellular connectivity, signaling parameters, mobility state of the device, etc. recorded in controlled environments.

Undoubtedly, such an approach becomes \textbf{unrealistic for 5G} as it is supposed to be characterized by wide diversity. For example, in the initial phase of deployment, 5G networks will coexist with the legacy 4G \ac{LTE} networks in the non-standalone mode of 5G. This, in turn, would result in a scenario where both low-power gNodeBs (gNBs) as well as legacy eNodeBs (eNBs) will exist in congruence. The situation's complexity will be further aggravated as the same operators may choose to support different technologies depending on the user's location, which will heavily influence the application's perceived throughput depending on the underlying technology. In addition, 5G also promises to support a wide range of devices, from low-power, short-range IoT devices to long-range mobile devices. This heterogeneity in different communication hardware often has a profound impact on the application's overall performance as factors like receiver sensitivity directly impacts the throughput, which, together with battery capacity, influences the final performance. Naturally, a model trained in restricted settings over a limited set of devices will fail to generalize, given this wide range of variations across different network technologies and devices.

A prudent solution in such a case can be \textbf{in-situ learning}, which allows the model to continuously capture the network states and device-specific parameters on the fly and retrain itself accordingly. However, for each device to train individual models, there will be a \textbf{requirement for an extensive training dataset} which might be \textbf{difficult to obtain}. Also, this model though more personalized, will not be robust to other devices or network settings if used. An idea to mitigate this problem of a dearth of training datasets and still train a \textbf{robust model} in a heterogeneous environment can be through \textbf{extracting knowledge collaboratively} obtained across different devices without violating their data privacy. Interestingly, we find that \textbf{Federated Learning} (FL) provides us with an avenue to train a global model across different devices over varying network technologies while maintaining the \textbf{data privacy}.

\begin{figure}
    \centering
    \includegraphics[width=0.60\columnwidth,keepaspectratio]{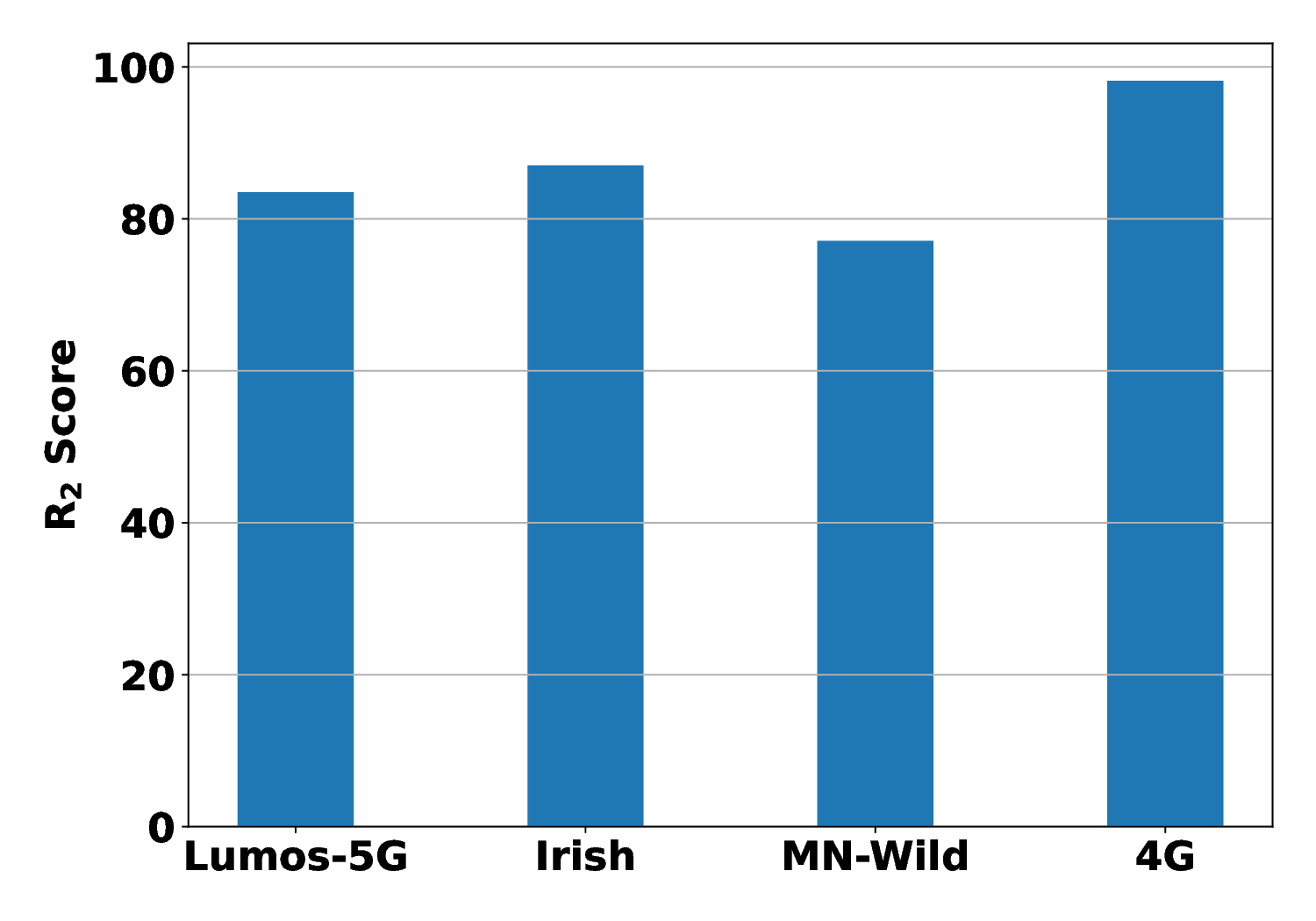}
    \caption{\textbf{Cross-technology Knowledge Transfer from 4G to 5G}. An effective way of performing throughput prediction over 5G can be done using a pre-trained model with a 4G dataset. We tested the proposed idea over publicly available 5G datasets along with an in-house collected 4G dataset. The results prove that pre-training with a 4G dataset gives more accurate throughput over 5G.}
    \label{fig:intro_plot}
\end{figure}
However, simply using the vanilla FL settings may fail to capture different layers of heterogeneity; for example, with 4G legacy devices still in use, a global model trained in FL setting over 5G may get colluded with the occasional data coming over 4G due to mobility. No wonder, when we perform a similar experiment, as shown in \figurename~\ref{fig:intro_plot}, we observe more \textbf{accurate predictions of throughput over 5G} if we use a \textbf{pre-trained model} that has been \textbf{bootstrapped} using datasets recorded \textbf{over 4G}. Understanding these benefits, in this paper, we propose \ourmethod{}, which uses the cross-technology knowledge extracted from 4G and applies it over 5G in collaborative FL-setup for accurate throughput predictions (Section~\ref{sec_methodology}). The key contributions of this paper are as follows.\\

\noindent\textbf{1. Collaborative \textit{in-situ} learning for network throughput prediction.} We show that a federated approach alleviates the problem of scarcity of data of 5G-enabled devices while making the model more robust and generalized across different hardware and network technologies.

\noindent\textbf{2. The 4G bootstrapping.} We show that the conventional approach of randomly initializing a model is ineffective. On the contrary, bootstrapping the initial model with diverse data obtained over 4G can actually boost the model for more accurate throughput predictions while making the system robust to data from legacy devices as well.

\noindent\textbf{3. Performance analysis over diverse setups and a PoC application.} From a thorough performance analysis over three different datasets collected from both real and simulated environments (Section \ref{sec:Dataset}), we show that \ourmethod{} can attain $>90\%$ mean $R_2$ score for throughput prediction in a multi-network multi-device environment, whereas the closest baseline has $<80\%$ mean $R_2$ score (Section \ref{sec_eval}). The efficacy of the throughput predictor has also been validated with a proof-of-concept (PoC) ABR streaming application.

\section{Related Work}\label{sec_Rel_work}
Throughput prediction has been studied extensively for WiFi Networks~\cite{KHAN2020}, Ethernet LAN~\cite{Jain2002} and for cellular networks~ \cite{Koutsonikolas2009,Yue2018, Zhang2019, Raca2019, Raca2019, Raca2020_1, Raca2020_1, Narayanan2020, Narayanan2021}, as summarized in Table \ref{tab:rel_work}. Time series forecasting-based throughput prediction methods have used statistical methods~\cite{raca2017back, Elsherbiny2020,lawal2020gmdh, Raca2019} such as Moving Average (MA), Auto-Regressive Moving Average (ARMA), or Auto-Regressive Integrated Moving Average (ARIMA), Exponential Smoothing,  Exponentially Weighted Moving Average (EWMA), etc. However, \cite{Raca2020, Raca2018_2, Mondal2020} have inferred that the cellular network throughput data
bears a non-trivial relation with the network parameters as well as the UE characteristics such as phone model, UE speed, etc.~\cite{Mondal2020}. Hence, time series models, like ARIMA and EWMA, have been found to perform not as well as their ML counterparts, as also shown in various other later works~\cite{Raca2020}. \textit{LinkForecast}~\cite{Yue2018}, one of the earliest learning-based throughput prediction algorithms, has used the popular \ac{RF} algorithm. The work shows that in addition to upper layer information, such as historical throughput, lower layer information like \ac{RSSI}, \ac{RSRP}, \ac{CQI}, etc.,  are integral towards accurate network throughput prediction.

In \cite{Raca2019}, authors have used historical throughput data to predict the average throughput over a finite future time window in a trace-driven controlled lab environment. The proposed throughput prediction of~\cite{Raca2019} has been used in \cite{Mondal2020} to improve the QoE performance and energy consumption of ABR video streaming algorithms. Various recent works~\cite{Schmid2019, Jiantao2020, Elsherbiny2020, Raca2020} have explored deep learning models for cellular throughput prediction. In \cite{Raca2020}, the authors have compared the performance of ML algorithms such as \ac{RF} Regressor and Support Vector Regressor with \ac{LSTM} for both raw data input and the summarization approach of \cite{Raca2019}. An important observation is that the  LSTM model has a shorter initialization and running time than the RF or SVR. \cite{Elsherbiny2020} compares the throughput prediction performance of ARIMA, K-nearest neighbor, Support Vector Regression, Ridge Factor Regression, \ac{RF} Regression, and LSTM. The results show that the \ac{RF} algorithm outperforms all the other algorithms. The authors attribute the improved performance of \ac{RF} to its generalization capability, which is made possible by introducing an additional level of randomness to the features. A location-independent throughput prediction approach using LSTM is proposed in~\cite{Schmid2019}. It shows that the selection of the hyperparameters, like the `lag' of LSTM, significantly affects the algorithm's performance. A combination of LSTM and CNN is also used in~\cite{Chauting2019} to propose an architecture called Spatio Temporal Cross-domain Neural Network (STCNet), which predicts the city-wide cellular network throughput using cross-domain data, such as base station information, Point-of-Interest (POI) distribution, and social activity level. Authors in~\cite{Parera2021} have also used transfer learning for predicting \ac{CQI} of UEs across different cities. Different cities serve as the source and target domains in this method. A combination of LSTM and Bayesian Fusion has been used in~\cite{Mei2020} to predict the bandwidth in different mobility scenarios, although it does not consider the impact of network parameters on throughput.

A recent work~\cite{Minovski2021} on 5G throughput prediction has compared the performance of \ac{RF}, Extreme Gradient Boosting Decision Trees (XGBoost), \ac{MLP}, and \ac{SVR}. Due to the limited availability of 5G commercial networks, the authors have first tested these algorithms on different 4G LTE network scenarios for validation. Subsequently, they have extended it to a 5G non-standalone network and then a 5G standalone network. \cite{Narayanan2020} treats the throughput prediction problem in 5G as both a classification and a regression problem using Gradient Boosting and Sequence to Sequence algorithms. The authors have also developed a 5G testbed for throughput measurement. Their analysis reflects that various factors govern the 5G throughput performance and differs significantly from its 3G or 4G counterparts. In \cite{Narayanan2021}, the authors have collected real-world 5G datasets based on two major 5G network service providers. Here authors have also studied the impact of throughput prediction in 5G video streaming applications. Authors in \cite{palit2023improving} have developed an \ac{ABR} video streaming application that improves the video playback quality and energy consumption by tuning the playback buffer to the state of the underlying 5G network, which is predicted using an \ac{LSTM} based model. In \cite{hassan2022vivisecting}, authors show how frequent handovers cause wild fluctuations in 5G throughput, which further degrades the application performance. They designed a handover prediction system to infer the correct network throughput needed for improving QoE for 5G video streaming applications.

\begin{table}[!t]
    \scriptsize
    \centering
    \caption{SOTA Approaches for throughput prediction}
    \begin{tabular}{|p{0.3cm}|p{1cm}|p{1.8cm}|p{3.8cm}|}
    \hline
    \textbf{Ref.} & \textbf{Network (3G/4G/5G or WiFi)} & \textbf{Method used} & \textbf{Geographical Area Covered} \\ \hline
    \cite{KHAN2020} & WiFi & MLP & Real home WiFi network \\ \hline
    \cite{Yue2018} & 4G/LTE & RF & Indoor and Outdoor 4G cellular data \\ \hline
    \cite{Raca2019} & 4G/LTE & RF & Static, pedestrian, bus, train, car, and highway \\ \hline
    \cite{Mondal2020} & 4G/LTE & RF & Major metropolitan and sub-urban cities of India\\ \hline
    \cite{Schmid2019} & 4G/LTE & RF, LSTM, SVR, DNN & Collected from two cities Amberg and Aschaffenburg\\ \hline
    \cite{Chauting2019} & GSM, CDMA, 4G/LTE & Transfer Learning (TL) based \ac{LSTM} & Across different regions in the city of Milan\\ \hline
    \cite{Mei2020} &  4G/LTE and HSPA & \ac{LSTM} RNN &  Long bandwidth traces on New York City MTA bus and subway.\\ \hline
    \cite{Minovski2021} & 4G/LTE, 5G &  \ac{RF} with XGBoost, SVR & Driving in urban, suburban, and rural areas, as well as tests in large crowded areas \\ \hline
    \cite{Narayanan2020} & 5G & DL based Gradient Boosting and Sequence to Sequence algorithms &  Urban areas covering roads, railroad crossings, restaurants, coffee shops, and outdoor recreational parks in Minneapolis \\ \hline
    \end{tabular}
    \label{tab:rel_work}
\end{table}

\textbf{Takeaways:} As mentioned above and summarized in Table~\ref{tab:rel_work}, all existing throughput prediction algorithms operate on centralized datasets, wherein they are trained in a central server using data from all connected \acp{UE}. However, reluctance to share proprietary network information by users and operators, as well as data privacy concerns, can make such centralized training non-lucrative. Furthermore, the diversification of 5G devices and user behavior restricts the applicability of a centralized throughput prediction algorithm. Therefore, these algorithms can perform well for the same trained datasets, but it is not fit for prediction in a multi-network, multi-device dataset. In the next section, we discuss data collection and a series of pilot experiments to highlight the limitations of these existing works.

\section{Data Acquisition}\label{sec:Dataset}
The primary objective of this paper is to develop a \textbf{robust} \textit{in-situ} model that can seamlessly predict the 5G cellular network throughput at any time instance from a set of network characteristics sensed by the end-device. Additionally, in this context, the term robustness means that the framework should be resilient to the fluctuations caused by the hardware components, the area-specific characteristics like population density, and variances in network characteristics introduced by different demographics and service providers' network architectures. However, a learning model for throughput prediction needs a vast amount of data to cover the diversity and scale of parameters inherent in cellular network modeling. To understand the heterogeneity across multi-network technologies, operators, device hardware, etc., we utilize three different datasets -- (a) in-house datasets collected over legacy 4G networks, (b) simulated 5G dataset using $\mathtt{ns3-mmwave}$~\cite{Mezzavilla2018} (5G-Simu), and (c) three publicly available 5G datasets (5G-Pub). The detail follows. 

\subsection{In-house Real 4G Dataset}\label{data}
The 4G network data used in this work has been collected considering the following primary factors: different user locations, mobility, Network Service Providers (NSP), and phone models. We have used two different smartphones for the setup -- a Micromax Canvas Infinity (M1) and a  Moto G5 (M2).
The throughput of these mobile phones has been recorded in buses, cars, and while walking in five different geographical locations in India (summarized in Table~\ref{tab:thpt_noise}). Cities 1 and 2 are large metropolitan areas, whereas City3, City4, and City5 are suburban areas.  All these cities have a high population density (a minimum of $2290$ persons/sq.km.). We have used the mobile Internet connections of three leading service providers in the country -- Airtel, Reliance JIO, and Vodafone Idea (Vi). The entire corpus of data traces has been collected over eleven months and amounts to more than 50 GB. Due to the difference in setup and session length, the size of individual datasets is different, as indicated in Table \ref{tab:thpt_noise}. Furthermore, as observed from the table, the throughput also shows variations in mean ($\mathbf{\overline{Y}}$) and standard deviation ($\mathbf{\delta_Y}$) based on the location, phone model, the service provider, and session length. 

The throughput profiling primarily focuses on file download applications with workloads of 6 MB, 100 MB, and 1 GB. An HTTP client-server program has been designed wherein the client runs on a rooted Android phone, and the server runs on an Amazon Web Server (AWS). Radio-related information has been collected using \textit{NetMonitorLite} App\footnote{\url{https://network-monitor-lite.soft112.com/} (Accessed: \today)}, and location and speed information have been captured using \textit{GPS Logger} App\footnote{\url{http://www.basicairdata.eu/projects/android/android-gps-logger/} (Accessed: \today)}. The throughput traces have been collected using $\mathtt{tcpdump}$ and analyzed using $\mathtt{Wireshark}$. 

The NetMonitorLite App records the Mobile Country Code (MCC), Mobile Network Code (MNC), Location Area Code (LAC), and, Cell ID (CID) of the associated base station. We have used these metrics to find the geographical coordinates of the \ac{BS} from \textbf{OpenCellId}\footnote{OpenCellId: \url{https://opencellid.org/} (Accessed: \today)}. The distance of \ac{UE} from the \ac{BS} was calculated from the \ac{UE} and \ac{BS} location coordinates. Finally, the following parameters have been captured in this dataset -- (a) \textbf{Radio Channel metrics}: \ac{RSSI}, data state, number of handover events, (b) \textbf{Location metrics}: UE geographical coordinates, UE speed, distance of UE from BS, and (c) \textbf{Downlink throughput}.
\begin{table}[t]
	\scriptsize
    \centering
    \caption{Noise variance in throughput data for various input-related factors and the size of the datasets}
    \begin{tabular}
    {|c|c|c|c|c|c|c|}
    \hline
    \textbf{Cities} & \begin{tabular}{@{}c@{}} \textbf{Phone} \\ \textbf{Model} \end{tabular} & \textbf{NSP} & \begin{tabular}{@{}c@{}} \textbf{Avg. Speed} \\ \textbf{(m/sec)} \end{tabular} & $\mathbf{\overline{Y}(Kbps)}$ & $\mathbf{\delta_Y (Kbps)}$ & \begin{tabular}{@{}c@{}} \textbf{No. of} \\ \textbf{entries} \end{tabular}\\ \hline
     City1 & M1 & Airtel & 4.23 & 0.77 & 1.09 & 1376\\ \hline
     City1  & M1 & JIO & 4.03 & 0.52 & 0.85 & 2273\\
     \hline
     City1 & M2 & JIO & 8.21 & 0.43 & 0.56 & 1518\\
     \hline
     City1 & M1 & Vi & 5.97 & 0.17 & 0.67 & 6201 \\
     \hline
     City2 & M2 & Airtel & 12.94 & 0.21 & 0.31 & 7932\\
     \hline
     City2 & M2 & Airtel & 5.07 & 0.31 & 0.81 & 7496\\
     \hline
     City3 & M1 & Airtel & 14.16 & 0.6 & 0.92 & 7354\\
     \hline
     City4 & M2 & JIO & 2.05 & 0.29 & 0.36 & 11008\\
     \hline
     City5 & M2 & JIO  &  0.06 & 0.69  & 0.024 & 965 \\
     \hline
    \end{tabular}
    \label{tab:thpt_noise}
\end{table}
\begin{table}
	\scriptsize
    \centering
    \caption{Noise variance in 5G throughput data for 10 \acp{UE} in the simulation}\label{tab:thpt_variance_5G}
     \begin{tabular}{|c|c|c|c|}
        \hline
        \textbf{User} & \textbf{Avg. Speed (m/sec)} & $\mathbf{\overline{Y} (Mbps)}$ & $\mathbf{\delta_Y (Mbps)}$\\ \hline
        1 & 12.8 & 8.53 & 7.48 \\ \hline
        2 & 16.9 & 3.15 & 3.97 \\ \hline
        3 & 12.5 & 8.36 & 8.39  \\ \hline
        4 & 18.2 & 3.31 & 3.87\\ \hline
        5 & 13.1 & 4.13 & 5.93 \\ \hline
        6 & 13.7 & 9.14 & 13.47 \\ \hline
        10 & 21.5 & 3.42 & 5.43 \\ \hline
    \end{tabular}
	\end{table}
\subsection{Synthetic 5G Dataset (5G-Simu)}\label{data_simulated}
The simulated data traces have been collected from a 5G mmWave network setup in ns3 network simulator\footnote{\url{https://www.nsnam.org/} (Accessed: \today)} with video streaming as the primary workload. The simulation scenario for simulating the behavior of a 5G \ac{UE}  has been set up using the $\mathtt{ns3-mmwave}$ module~\cite{Mezzavilla2018}. Some essential functions associated with the video download at the UE, such as ABR streaming algorithms, have been done in a Python setup. An HTTP-based DASH video streaming server that hosts the video segments has also been implemented. The details of the simulation setup are as follows.

\begin{figure}{%
	\centering
	\subfigure[]{%
	    \centering
		\label{fig:deployment_map}%
		\includegraphics[width=0.70\columnwidth]{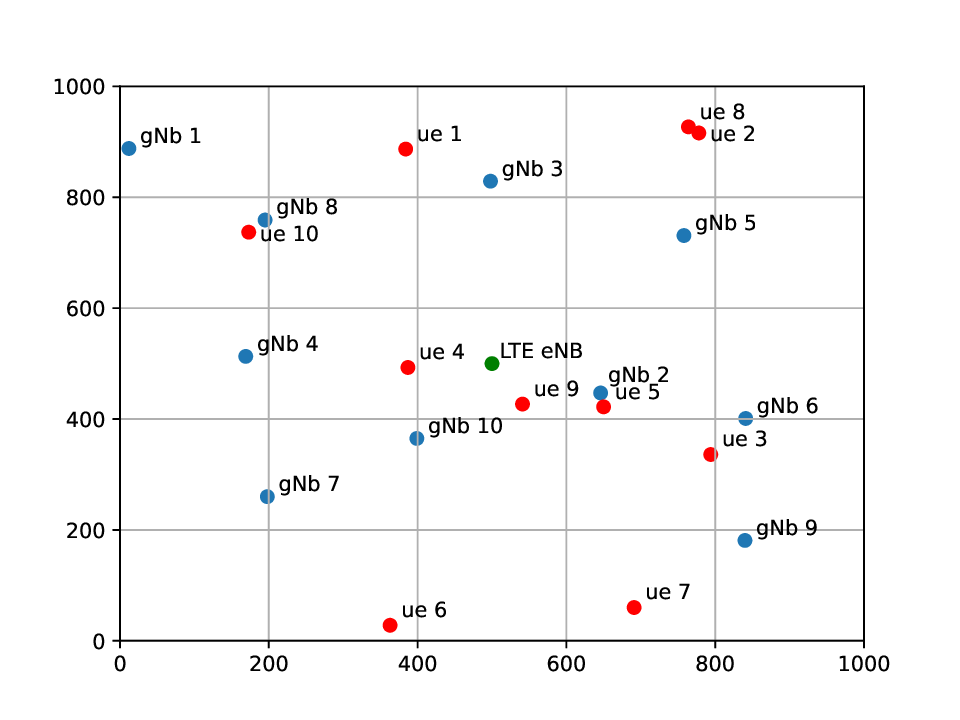}}
  
	\subfigure[]{%
	    \centering
		\label{fig:simulation_setup}%
		\includegraphics[width=0.70\columnwidth]{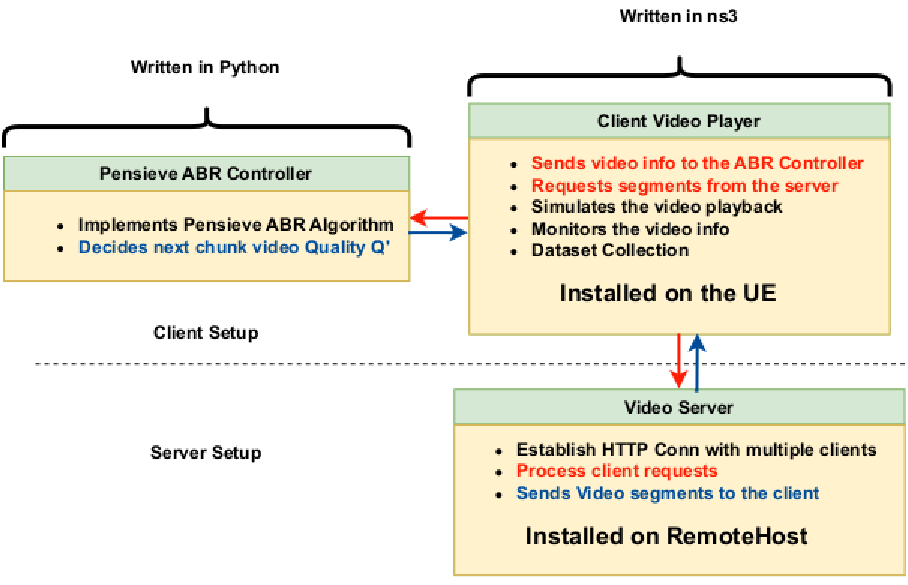}}
	\caption{\texttt{ns3-mmwave} simulation setup -- (a) deployment map, (b) simulation framework}
	\label{fig:5G_simulation_setup}}
\end{figure}
The simulation scenario, shown in \figurename~\ref{fig:deployment_map}, is a $1\times 1$ square kilometer area, inside which $10$ low-power 5G \acp{gNB} operating at the $28$ GHz frequency range are distributed uniformly. These \acp{gNB} govern the data communication between the \acp{UE}. A \ac{LTE} \ac{eNB} is located centrally in the simulation area. It operates in the sub-7GHz range and oversees the control channel communication between 5G gNBs. We have considered two different network loads corresponding to which $2$ and $10$  mobile \acp{UE} are distributed uniformly inside the simulation area. The mobility of this \acp{UE} has been modeled using the Random Walk mobility model. The average velocity of the \acp{UE} varies between $5$m/s to  $21$m/s (Details in Table~\ref{tab:thpt_variance_5G_2} and Table~\ref{tab:thpt_variance_5G}). We have deployed $10$ buildings inside the said simulation area to capture the effects of multipath fading and shadowing. Other simulation parameters are the same as in~\cite[Table I]{Mezzavilla2018}.

Each UE has an ongoing video streaming application running on its device, for which the throughput data has been collected. The \ac{DASH} video streaming has been implemented using a server-client system operating in the downlink. The DASH client application installed over the $\mathtt{ns3-mmwave}$ \ac{UE} fetches the video data rates from the video server installed over an $\mathtt{ns3-mmwave}$ remotehost. A Python DASH ABR proxy server connected to the video client application hosts the Pensieve ABR streaming algorithm~\cite{mao2017neural}. The role of this Python server is to decide the next video chunk quality based on the video information received from the UE. In \figurename~\ref{fig:simulation_setup}, we have shown our simulation framework. Each simulation scenario has been executed $10$ times with a random initial seed. From the simulation, the data recorded at each UE includes - (a) \textbf{network-related parameters}, such as the network throughput, \ac{RSSI}, SINR, \ac{MCS}, data state and the number of handovers, and (b) \textbf{UE parameters}, such as distance from the \acp{gNB}, device speed and energy consumption per bit. After the data collection phase, we further processed these data for playing the role of throughput prediction.  We have made both the 4G and 5G datasets publicly available in GitHub\footnote{\url{https://github.com/arghasen10/fedput-implement}}.

\noindent \textbf{Why simulated 5G?} 5G technology remains within the development phase, and we may have to wait for a short while before having an operational 5G network, as outlined by 3GPP standards, primarily in the middle and low-economy countries because of which it is still not accessible to many users across the globe. To understand the throughput behavior of next-generation cellular networks, many publicly available 5G datasets \cite{Raca2020_1, Narayanan2020, Narayanan2021} can help. Another alternative can be the simulated 5G dataset. The advantage of a simulated dataset is that it can be executed multiple times under the same network configuration seed and help regenerate the traces. One can tune the network configuration, and change the deployment scenario and topology for the data collection phase. Moreover, $1$s run with a sampling frequency of $100$ (or $10$ ms time interval for generating logs) can provide $100$ entries of throughput, thus producing adequate data for an accurate throughput prediction. 

\subsection{Publicly Available 5G Datasets (5G-Pub)}\label{public_data}
We have utilized three publicly available 5G datasets -- (1) \textbf{Lumos-5G} \cite{Narayanan2020}, (2) \textbf{Irish} \cite{Raca2020_1} and \textbf{MN-Wild}~\cite{Narayanan2021}. 

\subsubsection{Lumos-5G dataset} 
In \cite{Narayanan2020}, authors have conducted a measurement study of commercial 5G mmWave services in Minneapolis, MN, a major U.S. city, focusing on the downlink throughput as perceived by applications running on \ac{UE}. They have developed their Android application to log information such as \ac{UE}'s geographical coordinates, moving speed, compass directions, downlink throughput (reported using \textit{iperf 3.7}), radio type (4G/5G), signal strength (LTE -- RSRP, RSRQ, RSSI \& 5G -- SSRSRP, SSRSRQ, SSRSSI), handover events, etc. For data collection, they have selected three urban areas with mmWave 5G coverage -- (1) an outdoor four-way traffic intersection, (2) An indoor mall area inside Minneapolis-St. Paul (MSP) International Airport, (3) a 1300-meter loop near U.S. Bank Stadium covering roads, railroad crossings, restaurants, coffee shops, and outdoor recreational parks. With Verizon's 5G UW network, the measurement study is being conducted for $6$ months, using four Samsung Galaxy S10 5G smartphones.

\subsubsection{Irish dataset} 
In \cite{Raca2020_1}, the authors have generated 5G trace datasets collected from a major \textbf{Irish} mobile operator. They have considered two mobility patterns (static and car) and two user application patterns (video streaming and file download). The dataset consists of (1) channel-related metrics such as signal strength (RSRP, RSRQ, SNR, CQI), neighboring cell RSRP, RSRQ, (2) context-related metrics (e.g., GPS of the device, device velocity), (3) cell-related metrics such as eNBs ID, and (4) throughput information for both uplink and downlink. For obtaining the dataset, they have used \textit{G-NetTrack Pro}, an Android network monitoring application. The dataset
contains $83$ traces, with a total duration of $3142$ minutes, using a Samsung S10 5G Android device.

\subsubsection{MN-Wild}
In \cite{Narayanan2021}, authors have carried out an in-depth measurement study of the performance, power consumption, and application QoE of commercial 5G networks in the wild. They have examined different 5G carriers (Verizon and T-Mobile), deployment schemes (Non-Standalone, NSA vs. Standalone, SA), radio bands (mmWave and sub-6-GHz), Radio Resource Control state transitions for power modeling, mobility patterns (stationary, walking, driving), client devices such as Samsung Galaxy S20 Ultra 5G (S20U) and Samsung Galaxy S10 5G (S10), and upper-layer applications (file download, video streaming, and web browsing). The entire measurement studies were conducted in two US cities (Minneapolis, MN and Ann Arbor, MI), where both carriers have deployed 5G services. The dataset is publicly available in GitHub\footnote{\url{https://github.com/SIGCOMM21-5G/artifact} (Accessed: \today)}. However, T-Mobile works under low-band 5G, deployed in both SA and NSA modes. Since our focus is more on the mmWave characteristics, we have selected only the dataset corresponding to the default mode of Verizon carrier. Among the two cities' data, we find that the dataset corresponding to Minneapolis city (MN) contains throughput and other important UE information, such as the speed of the UE, which is not available for the Ann Arbor (MI) dataset. Therefore here, we have selected the MN dataset. We name it as MN-Wild. The feature space for all these datasets is summarized in Table~\ref{tab:feature_space}.



\begin{table}
\scriptsize
\centering
    \caption{Noise variance in 5G throughput data for two \acp{UE} in the simulation}\label{tab:thpt_variance_5G_2}
     \begin{tabular}{|c|c|c|c|}
    \hline
    \textbf{User} & \textbf{Avg. Speed (m/sec)} & $\mathbf{\overline{Y} (Mbps)}$ & $\mathbf{\delta_Y (Mbps)}$ \\ \hline
    1 & 5.86 & 18.53 & 17.48\\ \hline
    2 & 14.1 & 14.114 & 12.57\\ \hline
    \end{tabular}
\end{table}

\begin{table}[t]
	\scriptsize
    \centering
    \caption{Raw features available across different datasets}
    \label{tab:feature_space}
    \begin{tabular}{|c|c|c|c|c|c|}
    \hline
        \begin{tabular}{@{}c@{}} \textbf{Feature} \\ \textbf{Name} \end{tabular} & \textbf{4G} & \textbf{Lumos-5G} & \textbf{Irish} & \textbf{MN-Wild} & \begin{tabular}{@{}c@{}} \textbf{Synthetic} \\ \textbf{5G} \end{tabular} \\ \hline
        Timestamp & $\pmb{\checkmark}$ & $\pmb{\checkmark}$ & $\pmb{\checkmark}$ & \pmb{\checkmark} & $\pmb{\checkmark}$ \\ \hline
        Lat, Long & \checkmark & \checkmark & \checkmark & \checkmark & X \\ \hline
        Radio type (4G/5G) & $\pmb{\checkmark}$ & $\pmb{\checkmark}$ & $\pmb{\checkmark}$ & $\pmb{\checkmark}$ & $\pmb{\checkmark}$ \\ \hline
        Speed & $\pmb{\checkmark}$ & $\pmb{\checkmark}$ & $\pmb{\checkmark}$ & \pmb{\checkmark}& $\pmb{\checkmark}$ \\ \hline
        Operator Name & \checkmark & X & X & \checkmark& X \\ \hline
        Horizontal Handover & $\pmb{\checkmark}$ & $\pmb{\checkmark}$ & $\pmb{\checkmark}$ & $\pmb{\checkmark}$ & $\pmb{\checkmark}$ \\ \hline
        Vertical Handover & \checkmark & \checkmark & X &X & X \\ \hline
        \begin{tabular}{@{}c@{}} Signal Strength \\ (RSRP/RSRQ/RSSI) \end{tabular} & $\pmb{\checkmark}$ & $\pmb{\checkmark}$ & $\pmb{\checkmark}$ & \pmb{\checkmark}& $\pmb{\checkmark}$ \\ \hline
        SNR & X & \checkmark & \checkmark & \checkmark & \checkmark \\ \hline
        CQI & X & X & \checkmark & X & \checkmark \\ \hline
        Throughput & $\pmb{\checkmark}$ & $\pmb{\checkmark}$ & $\pmb{\checkmark}$ &\pmb{\checkmark} & $\pmb{\checkmark}$ \\ \hline
        Data State & \checkmark & X & \checkmark & X & \checkmark \\ \hline
        Distance from cell & \checkmark & X & X & X & \checkmark \\ \hline
    \end{tabular}
\end{table}
\normalsize
\section{Pilot Study}\label{sec_motivation}
Before diving into the design of \ourmethod{} pipeline, we first perform an in-depth analysis of some of the existing state-of-the-art algorithms for multi-network (across 4G and 5G systems) and multi-device throughput prediction. The detail follows. 

\subsection{Issues with SOTA Throughput Predictors}
As we discussed earlier, recent approaches for 5G throughput prediction have primarily used two different state-of-the-art models -- \ac{LSTM}~\cite{Raca2020} and \ac{RF}~\cite{Minovski2021}. To analyze their performance over multi-network and multi-device predictions, we perform three separate experiments -- (a) train the model using one dataset from 5G-Pub (see Section~\ref{public_data}), and then test with the held-out data from the same or a different dataset from 5G-Pub, (b) train the model using a centrally mixed 4G \& 5G-Pub dataset, and then train using the held-out data from 5G-Pub, and (c) train the models using the 4G dataset and then test using 5G-Pub or the held out 4G data. 

\begin{table}[!htb]
    \scriptsize
    \centering
    \caption{Percentage $R_2$ score for throughput prediction over 5G public datasets for training using SOTA approaches}
    \label{tab:thptPredAcrossData2}
    \begin{tabular}{|p{1.1cm}|p{0.3cm}|p{0.4cm}|p{0.4cm}|p{0.4cm}|p{0.3cm}|p{0.4cm}|p{0.4cm}|p{0.4cm}|p{0.3cm}|} \hline
    \textbf{Train set} & \multicolumn{3}{|c|}{\textbf{Lumos-5G (L)}} & \multicolumn{3}{|c|}{\textbf{Irish (I)}} & \multicolumn{3}{|c|}{\textbf{MN-Wild (M)}} \\ \hline
    \textbf{Test set} & \textbf{L} & \textbf{I} & \textbf{M} & \textbf{L} & \textbf{I} & \textbf{M} & \textbf{L} & \textbf{I} & \textbf{M} \\ \hline 
    \textbf{LSTM}~\cite{Raca2020} & \cellcolor{blue!25}95.1 & 58.2 & 57.7 & 46.5 & \cellcolor{blue!25}94.9 & 77.5 & 68.2 & 21.5 & \cellcolor{blue!25}96.8 \\ \hline
    \textbf{RF}~\cite{Minovski2021} & 81.4 & $<0$ & $<0$ & $<0$ & 18.7 & $<0$ & $<0$ & $<0$ & 90.6\\ \hline
    \end{tabular}
\end{table}
For the first experiment, we consider the three datasets from 5G-Pub -- Lumos-5G (L), Iris (I), and MM-Wild (M). We train the two models (RF and LSTM) using 70\% data from one of these three datasets and then test using either the remaining 30\% held-out data from the same dataset or one of the two other datasets. We use the features which are common to all three datasets (Table~\ref{tab:feature_space}). To evaluate the performance of the models, we use the percentage $R_2$ score. The results are summarized in Table~\ref{tab:thptPredAcrossData2}. The table indicates that for both models, the maximum performance is achieved when the models are trained and tested over the same dataset, emphasizing performance benefits for training and testing over the same data collection environment. However, for cross-environment testing, we observe a significant drop in the $R_2$ score. This clearly indicates that the \textbf{state-of-the-art throughput predictors} are \textbf{biased towards the environment} from where the data is collected.

\begin{table}[!htb]
	\scriptsize
    \centering
    \caption{Percentage $R_2$ score for throughput prediction for 4G and the merged datasets for training using SOTA approaches}
    \label{tab:thptPredAcrossData3}
    \begin{tabular}{|c|c|c|c|c|c|c|c|} \hline
    \textbf{Train Dataset} & \multicolumn{4}{|c|}{\textbf{4G}} & \multicolumn{3}{|c|}{\textbf{Merged}}\\ \hline
    \textbf{Test Dataset} & \textbf{L} & \textbf{I} & \textbf{M} & \textbf{4G} & \textbf{L} & \textbf{I} & \textbf{M} \\ \hline 
    \textbf{LSTM}~\cite{Raca2020} & \cellcolor{blue!25}83.5 & \cellcolor{blue!25}87.03 & \cellcolor{blue!25}77.1 & \cellcolor{blue!25}98.18 & 57.81 & 60.39 & 51.77 \\ \hline
    \textbf{RF}~\cite{Minovski2021} & 12 & 18.2 & 8.96 & 64.3 & $<0$ & $<0$ & $<0$ \\ \hline
    \end{tabular}
\end{table}

Next, we train the two models over a merged dataset where we mix the data ($70\%$ train data) from all four datasets, i.e., in house 4G dataset and the three publicly available 5G datasets, and then test the model's performance over the remaining $30\%$ held-out data from the three 5G datasets. We observe that prediction models perform miserably poor over the test data (see Table~\ref{tab:thptPredAcrossData3}). We notice that the diversity in the mixed data confuses the model when trained with the merged dataset. The reason is the differences in responsiveness for different UE hardware models and the variety in network configurations across operators and technologies. Thus, the prediction model cannot capture such diversity, particularly the device and technology diversities over a limited dataset. The \ac{RF} model results in a negative $R_2$ score, indicating that the model learns an opposite behavior. 

Finally, we perform cross-technology throughput prediction. For this purpose, we train the model with the 4G dataset and test it over the different 5G-Pub datasets. This time also we observe a fall in the prediction accuracy compared to the experiment when the prediction was made for the same dataset (see Table~\ref{tab:thptPredAcrossData3}). It is also evident that these state-of-the-art throughput prediction approaches are not suitable for multi-network scenarios. However, one interesting observation from this analysis is that the models perform a little better when trained on the 4G dataset and tested over a 5G dataset in comparison to multi-network 5G datasets. The 4G dataset, being more robust and stable than the 5G ones, can demonstrate specific patterns in the throughput, which the 5G models can use for bootstrapping. In the next set of experiments, we explore this further. 

\subsection{Bootstrapping with 4G Data}
\begin{figure}[!htb]%
	\centering
	\subfigure[]{%
	    \centering
		\label{fig:spearman_corr}%
		\includegraphics[width=0.45\columnwidth,keepaspectratio]{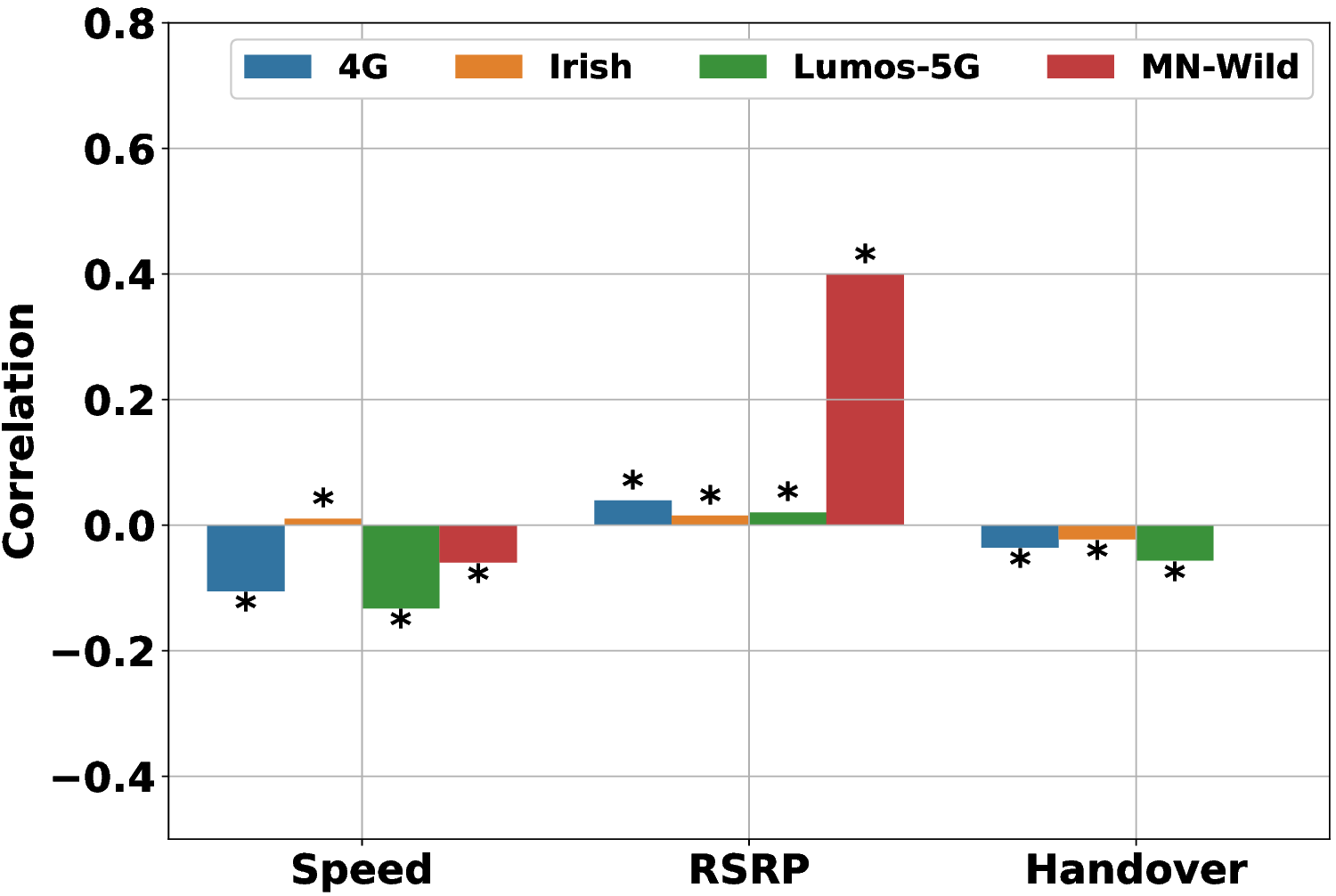}}\hfil
	\subfigure[]{%
	    \centering
		\label{fig:distribution_RSSI}%
		\includegraphics[width=0.45\columnwidth,keepaspectratio]{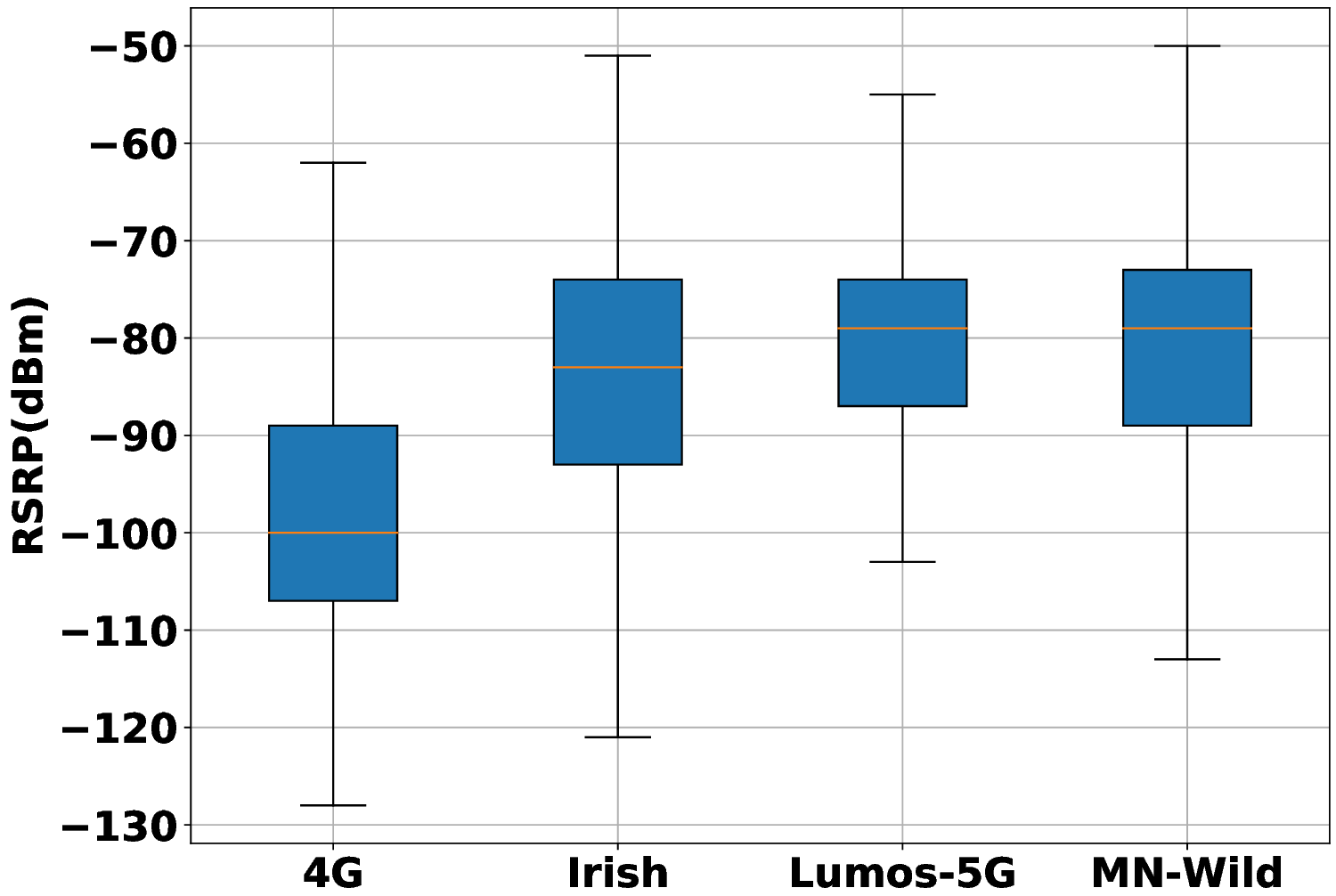}}
	\caption{Analysis of features across 4G and 5G (a) Spearman correlation of the input features with throughput for 4G dataset and different 5G datasets like \textbf{Irish}, \textbf{Lumos-5G}, and \textbf{MN-Wild} datasets (* - indicates $p$-value$<0.05$) and (b) variation in the RSRP distribution}
	\label{fig:correlation}
\end{figure}
To answer the above question, we start with pilot experiments considering the analysis of $3$ primary features used in the existing throughput predictors. These features are -- Speed of the \acs{UE}, \ac{RSRP}, and the number of handovers experienced by the \acp{UE}, as recorded in the publicly available datasets like Irish, Lumos-5G \& MN-Wild, and the collected 4G dataset. To begin with, we first observe the Spearman correlation of each of these features with the target variable, which is the downlink throughput. As shown in \figurename~\ref{fig:spearman_corr}, we observe strong consistency between the primary features regarding their impact on the overall throughput. Interestingly, this consistency is also present across the technologies, allowing us to monitor, exploit, and analyze the existing data available from legacy 4G devices to develop a more generalized model for throughput prediction in 5G. However, amidst all these opportunities, there are specific challenges.

\begin{figure*}[!htb]%
	\centering
	\subfigure[]{%
	    \centering
	    \label{fig:device_tsne}%
		\includegraphics[width=0.3\textwidth,keepaspectratio]{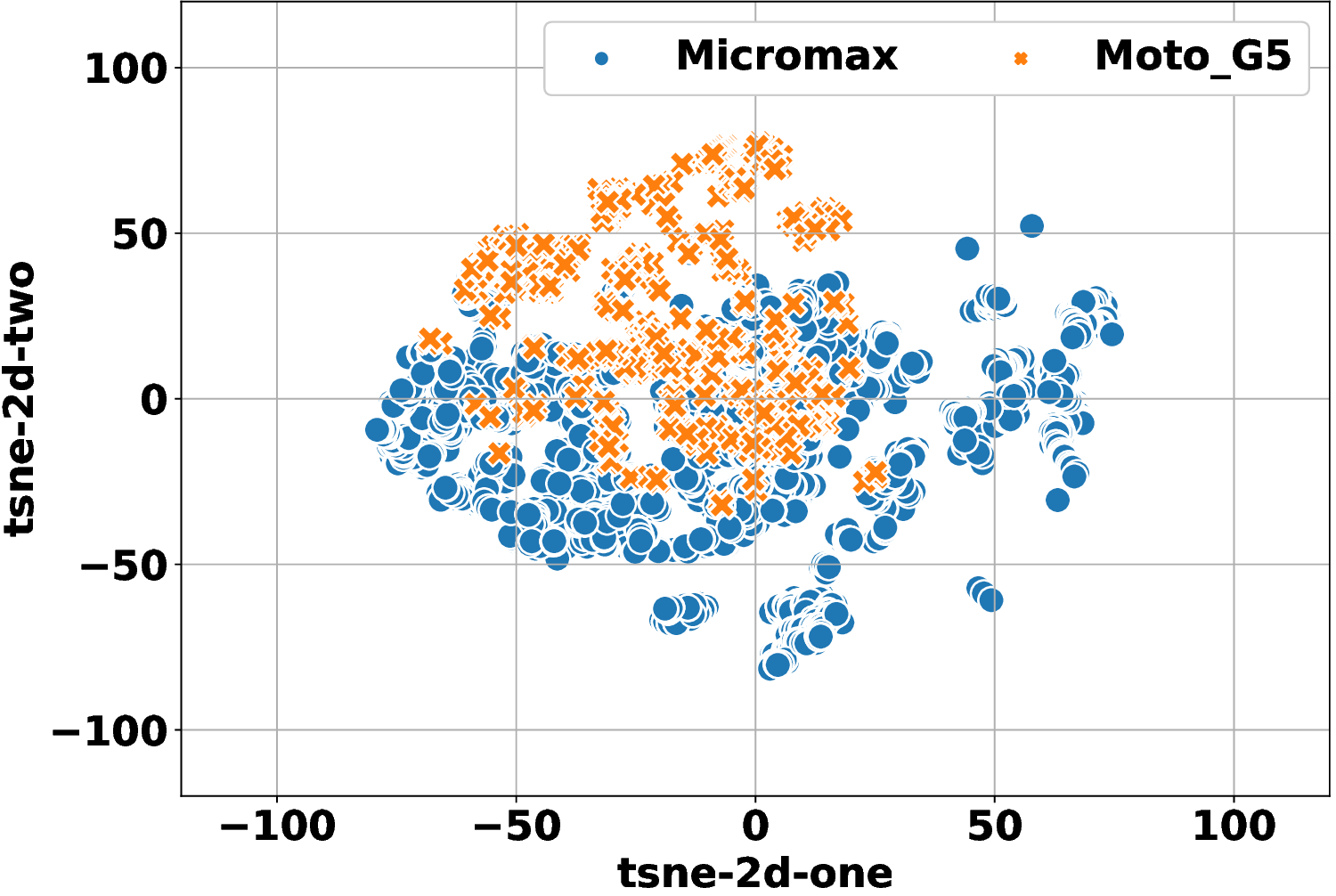}}\hfil
	\subfigure[]{%
	    \centering
		\label{fig:operator_tsne}%
		\includegraphics[width=0.3\textwidth,keepaspectratio]{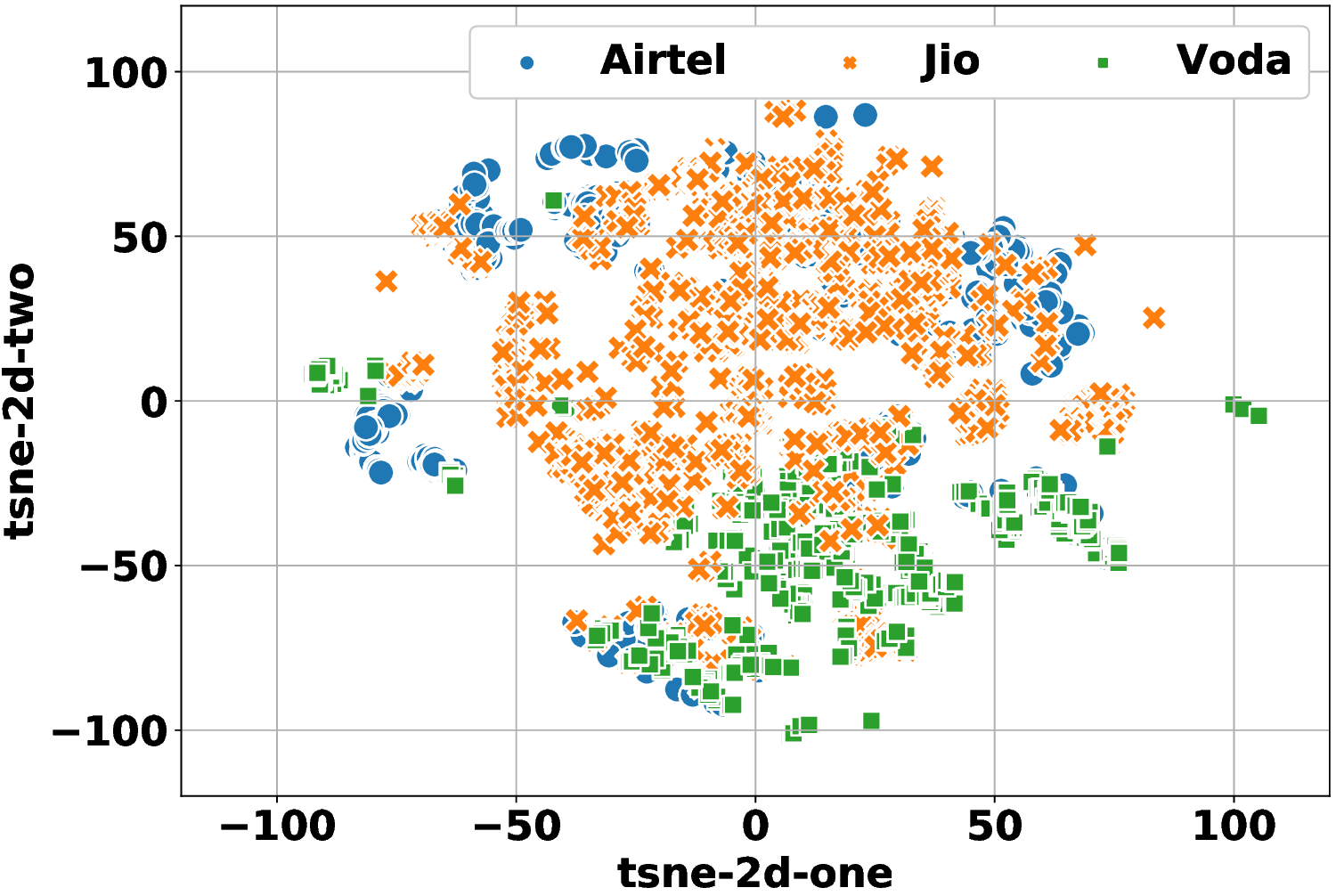}}\hfil
	\subfigure[]{%
	    \centering
		\label{fig:5G_operator_tsne}%
		\includegraphics[width=0.3\textwidth,keepaspectratio]{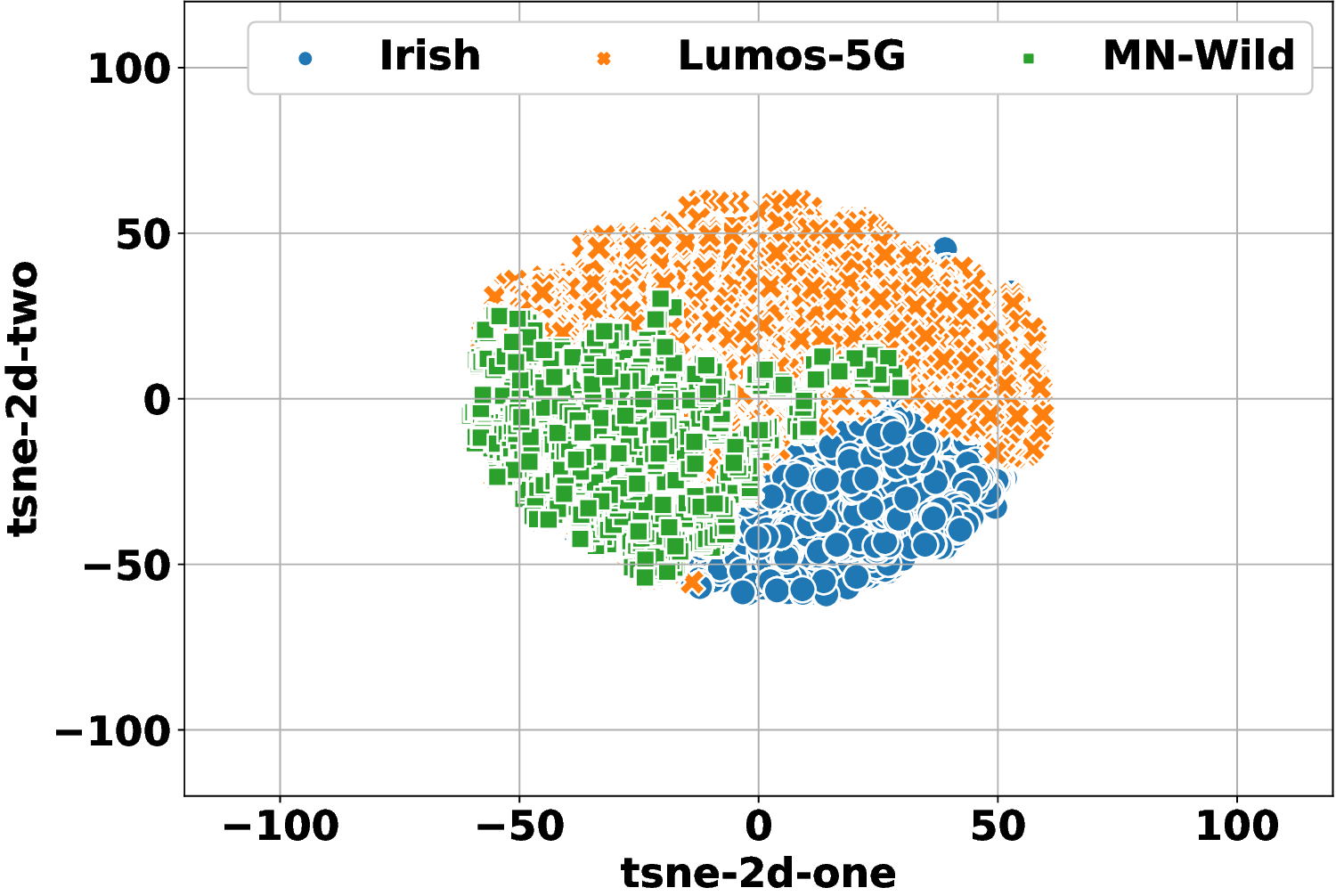}}\hfil
	\caption{Analysis of heterogeneity in the features across 4G and 5G technologies using t-SNE (a) device heterogeneity in 4G, (b) operator heterogeneity in 4G, (c) heterogeneity in publicly available 5G datasets}
	\label{fig:feature_Heterogeneity}
\end{figure*}

The first challenge we observe is the usual shift of data distribution across technologies like 4G to 5G. For example, the typical median RSRP values for 5G are slightly higher than that of the RSRP values observed in the 4G dataset. RSRP value primarily depends on the deployment scenarios; for example, if the transmission power of the co-channel neighbor base stations is high, then the RSRP value may degrade due to high interference. It also depends on the received signal strength of the UE, which is not similar for 4G and 5G. Thus, we have a significant variation in the distribution of RSRP values for the 4G and 5G datasets, as shown in \figurename~\ref{fig:distribution_RSSI}. Such domain shifts and differences in the range of values can impact the generalization of any model. Thus, a straightforward global model trained on 4G data might not identify the variations correctly, leading to erroneous throughput predictions, similar to the case with the centralized dataset. Secondly, a deeper analysis of the collected 4G dataset reveals that the feature space varies significantly with the underlying hardware and the service provider (see \figurename~\ref{fig:feature_Heterogeneity}). Considering the consistency of feature relations with throughput across 4G and 5G, we anticipate a similar presence of heterogeneity in 5G data. Such heterogeneities indicate local device and operator-specific patterns, which might reduce the performance of any standalone global model initialized and trained at some fixed instance in time. 
\section{Brief Overview of the Proposed Solution}
Say a mobile device $\mathcal{M}$ from a geographical area $\mathcal{G}$ is connected to an available operator $\mathcal{X}$ running an application $\mathcal{A}$. The primary objective of this paper is to develop a \textbf{robust} \textit{in-situ} model that can seamlessly predict the 5G cellular network throughput $Y_t = f(S_t)$ at any time instance $t$ from a set of network characteristics $S_t$ sensed by the end-device. Additionally, in this context, the term robustness means that the framework should be resilient to the fluctuations caused by the hardware components of $\mathcal{M}$, the area-specific characteristics like population density, and variances in network characteristics introduced by different operators. 

\begin{figure}[!t]%
	\centering
        \includegraphics[width=0.60\columnwidth]{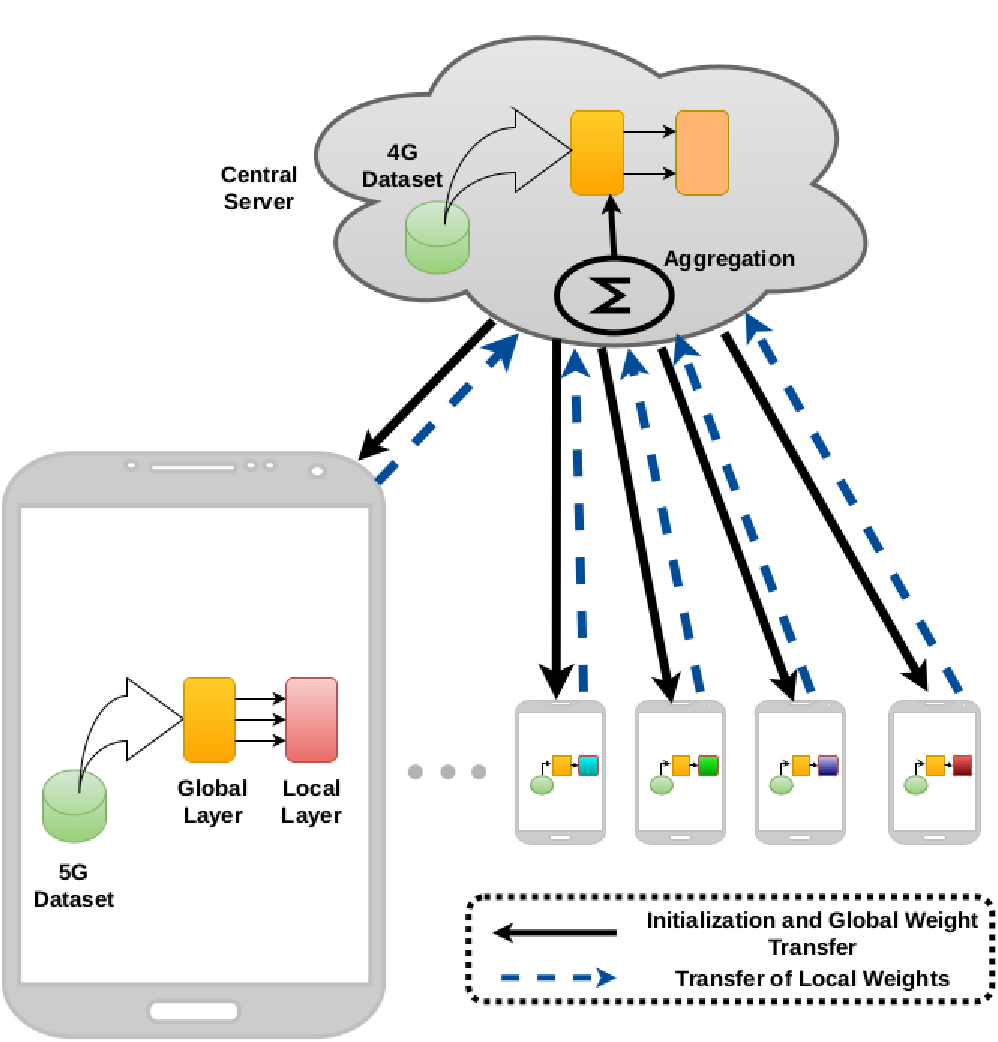}
	\caption{System framework of \ourmethod{}}
	\label{fig:Model_plot}
\end{figure}

Given the heterogeneity of 5G deployments in terms of the service providers or underlying communication hardware of the \ac{UE}s, a central learning model for throughput prediction cannot capture device-specific and network-specific parameters. A possible solution can be personalized \textbf{in-situ learning}, which demands an extensive training dataset that is difficult to obtain due to the nascent deployment of 5G primarily in the middle and low-economy countries. We find \ac{FL} based approach can mitigate this problem of a dearth of the training datasets and still train a robust model in a heterogeneous setup via extracting knowledge collaboratively across different devices~\cite{Konecny2016, konevcny2015federated, Jakub2016}. Understanding the challenges and opportunities from the pilot experiments, we develop \ourmethod{} as shown in \figurename~\ref{fig:Model_plot}. It consists of two main components, (i) A central server, and (ii) various 5G end-devices working in a federated setup. The central server hosts a recurrent neural network-based global model. The model consists of two layers of LSTM cells (128 units each) followed by a fully connected layer. After every LSTM layer, a dropout of 0.2 is added as a form of regularization. The global model is summarized in Table~\ref{tab:model_summary}. The two layers correspond to the two parts of the model - the first \ac{LSTM} layer is the global part $M_G$, whose weights can be updated globally for all users, and the second \ac{LSTM} layer $M_L$ is specific to individual users. The central server trains this global model with the preprocessed 4G dataset, due to its superior performance in predicting network throughput with 5G test datasets (summarized in Table~\ref{tab:thptPredAcrossData3}). It then shares this global model, initialized with the 4G data, with all the 5G end-devices. Each end-device then fine-tunes the local model with its locally collected preprocessed dataset. 
\section{Designing \ourmethod{}}\label{sec_methodology}

The next set of tasks that we carry out through a series of preprocessing steps includes -- (a) noise removal and (b) data formatting.
\begin{table}[ht]
	\centering
 \scriptsize
	\caption{Model architecture details}
	\label{tab:model_summary}
	\begin{tabular}{|c|c|c|c|c|c|} \hline
		\textbf{Layer} & \textbf{LSTM1} & \textbf{Dropout1} & \textbf{LSTM2} & \textbf{Dropout2} & \textbf{Desnse} \\ \hline
		\textbf{Output} & (5, 128) & (5, 128) & (128) & (128) & (1)\\ \hline
		\textbf{Param} & 69120 & 0 & 131584 & 0 & 129 \\ \hline
	\end{tabular}
\end{table}
\subsection{Preprocessing}
In typical 5G networks, there are several hidden parameters that have a direct impact on the throughput. However, all these hidden latent factors cannot be measured straight away. For example, the load condition of the base stations cannot be measured at the user end. Similarly, the users cannot quantify the effect of resource scheduling algorithms without input from the service providers. In this work, we treat the effect of these latent parameters as noise. So in the preprocessing step, the dataset is first passed through a Gaussian filter to remove the effect of noise.

Once the preprocessing steps are performed on the dataset, in the next step, we format the data into timesteps so that it can be used to exploit the time-series nature of the dataset for prediction using the designed model discussed in the following subsection. We format the data as follows. At every time step $i$,  we create $(\Vec{\psi_{X}^i}, \Vec{\psi_{Y}^i})$ - two data matrices from the filtered dataset for the domain 4G or 5G. Here, $X = \{X_1, X_2, ..., X_n\}$ corresponds to the set of input features in the filtered dataset, and $Y$ corresponds to the target throughput.  $\Vec{\psi_{X}^i}$ represents the matrix of network parameters and location-related input features from timestep $i-H$ to $i$ for a historical time window $H$, i.e., 
\begin{equation}\label{eq:createsample}
	{\Vec{\psi_{X}^i}} = \begin{pmatrix}
		X_{1}^{(i-H)} & X_{2}^{(i-H)} & ... & X_{n}^{(i-H)}\\
		X_{1}^{(i-H-1)} & X_{2}^{(i-H-1)} & ... & X_{n}^{(i-H-1)}\\
		\vdots & \vdots & \vdots & \vdots\\
		X_{1}^{(i-1)} & X_{2}^{(i-1)} & ... & X_{n}^{(i-1)} \end{pmatrix}.
\end{equation}
$\Vec{\psi_{Y}^i}$ is the vector of downlink throughput from $i-H$ seconds to $i-1$ seconds, represented as., 
\begin{equation}\label{eq:createsample_thpt}
	\Vec{\psi_{Y}^i} = [Y^{(i-H)} \ Y^{(i-H-1)} \ \cdots Y^{(i-1)}],
\end{equation}


\subsection{The Global Model}
Traditionally in a federated setup, the model deployment starts with an initialization step that includes setting up the global model. In this paper, we exploit the existing legacy 4G technology to obtain a bootstrapping dataset during the global model's initialization. The first step in such cross-technology setups is defining a judicious set of features that can then be used to initialize and train the global model. 
\subsubsection{Defining the Feature Space}
Notably, the user throughput in cellular networks depends on the user location, distance from the connected base stations, user speed, and network-related parameters, such as \ac{RSSI}, \ac{RSRP}, \ac{MCS}, data state, number of handovers, the technology of associated and neighboring base stations, among others. In this paper, we first select a standard set of available features both in the bootstrapping 4G dataset and the local 5G datasets. This set of features chosen from these two datasets includes the distance from the connected base stations, user speed, \ac{RSRP}, and the number of handovers. 

Once these features are defined, we next start developing and bootstrapping the global model as follows.
\subsubsection{Initialization and Training of Global Model}
The throughput prediction algorithm aims to predict the cellular network throughput over a window of $W$ seconds into the future based on network-related and location parameters, and the downlink throughput of the previous `$H$' seconds. To train the global-\ac{LSTM} network in this phase, we have used the in-house 4G dataset. The details of the dataset are provided in Section~\ref{data}. 
Here the objective is to learn the weights $\theta_{4G}$ for both the \ac{LSTM} layers.

The LSTM network in the global domain is trained using the 4G dataset, $\mathcal{D}_{4G}$ 
, specifically on $\Vec{\psi}_{X, 4G}^i, \Vec{\psi}_{Y, 4G}^i$ (see equation~\ref{eq:createsample}, \ref{eq:createsample_thpt}) to learn the weights $\theta_{4G}$ by minimizing the loss between the predicted average throughput $\hat{Y}_{4G}^{(i+W)}$ and the actual average throughput $\overline{Y}_{4G}^{(i+W)}$ (also known as mean average error). The predicted throughput $\hat{Y}_{4G}^{(i+W)}$ is obtained as,
\begin{align}\vspace*{-0.5cm}\label{eq:ypred}
\hat{Y}_{4G}^{(i+W)} =  \mathcal{F}((\mathbf{\psi}_{X, 4G}^i,\psi_{Y, 4G}^i),\overline{Y}_{4G}^{(i+W)}, \mathbf{\theta_{4G}}),
\end{align}
and the actual average throughput $\overline{Y}_{4G}^{(i+W)}$ is given by:
\begin{equation}\label{eq:thpt_avg_actual}
\overline{Y}_{4G}^{(i+W)} = \frac{1}{W}\sum_{j=i}^{i+W}Y_{4G}^{j}.
\end{equation}
Here $\mathcal{F}$ is the predictive function to be learned. Learning $\mathcal{F}$ is equivalent to learning the weights $\theta_{4G}$.

\subsection{Local Training}
Once the entire global model is trained and initialized on the 4G bootstrapping dataset, our prediction algorithm moves to the next phase, retraining the LSTM for the actual 5G mmWave connecting the end-devices. In this phase, the training takes place as shown in \figurename~\ref{fig:Model_plot}. The central aggregator, which hosts the LSTM with two layers $M_G$ and $M_L$, is connected to all the users. There are $U$ datasets $\{ D_1 , D_2 , ... , D_U\}$ belonging to  `$U$'  different users\footnote{In this paper, we use the words user and end-device interchangeably.} connected using the 5G mmWave network. 


Before the first iteration, the end-device downloads the current LSTM model available at the central server with the weights $\theta_{4G} = \{\theta_G, \theta_u\}$ corresponding to the layers $M_G$ and $M_L$.
The local model training then starts and takes place in epochs.
It takes as input - a) the users' datasets $D_u$, b) the global and user-specific model weights $\theta_G$ and $\theta_u$, c) the number of users $U$, and d) a set of parameters $\{\mathcal{P}\} = \{H,W,\sigma\}$, where $H$ is  history window length, $W$ is the prediction window length, and $\sigma$ is the standard deviation of Gaussian filter.

Once the iteration begins, the end-device assigns the global ($M_G$) and local layer ($M_L$) weights as $\theta_G,\theta_u$, respectively.
At this point at each user, the weight of the local layer ($\theta_u$) is the same. The local dataset is first preprocessed. The filtered dataset, which is suitably formatted using (Eq. \ref{eq:createsample}-\ref{eq:createsample_thpt}), is used for retraining the model locally. Mathematically, the local samples $\{\Vec{\psi}^i_{X,5G}, \Vec{\psi}_{Y,5G}^i \}$ are created.
The LSTM model is subsequently trained to learn the new weights $\theta_G^{new}$ and $\theta_u^{new}$ corresponding to user $u$, such that
\begin{equation}\label{eq:weight_tune}
\theta_{G}^{new} , \theta_u^{new}  = \argmin{\theta_G , \theta_u} \text{loss}(\hat{Y}_{5G}^{(i+W)}, \overline{Y}_{5G}^{(i+W)}).
\end{equation}

At any time step `$i$' the average throughput of user $u$ over a future time window of $W$ seconds is predicted as:
\begin{equation}
\label{eq:ypred_avg}
\hat{Y}_{5G}^{(i+W)} = \mathcal{F}((\Vec{\psi}^i_{X,5G}, \Vec{\psi}_{Y,5G}^i, {\theta_G} , {\theta_u}),
\end{equation}
As before, here, $\Vec{\psi}^i_{X,5G}$ are the network or location-related features, and $\Vec{\psi}_{Y,5G}^i$ the corresponding throughput data for the user. These represent the values from the filtered dataset.
Here $\overline{Y}_{u}^{(i+W)}$ gives the actual average throughput over the future $W$ seconds is given by
\begin{equation}\label{eq:thpt_avg}
\overline{Y}_{5G}^{(i+W)} = \frac{1}{W}\sum_{j=i}^{i+W}Y_{5G}^{j}
\end{equation}
\subsection{Aggregation and Local Prediction}
After the local retraining using the 5G mmWave dataset, the retrained local weights $\{\theta_u, \forall u\}$  are saved locally.
The new global weights generated by each user are sent to the server, where they are averaged using federated averaging
~\cite{Smith2017} to get the current global weight $\theta_G^\mathrm{new}$. In the next federated iteration, the global layer $M_G$ for each local model at each user $u$ is initiated with the new global weight, the aggregate of all the weights for the layer $M_G$ across all the end-devices $u$ obtained in the previous iteration.

During inferencing, each user instantiates its LSTM  model with the current global  $\theta_G$ and the current local weight $\theta_u$. The test dataset $D_{test}$ containing the historical network parameters and throughput information is filtered using preprocessing steps and then fed to the inferencing engine for the throughput prediction.
\section{Evaluation}\label{sec_eval}
We evaluate the performance of \ourmethod{} with respect to baseline throughput prediction algorithms widely used in the literature. Then we analyze the effect of \ourmethod{} on the popular \ac{ABR} video streaming applications. The details follow. 

\subsection{Implementation Details of \ourmethod{}}
\ourmethod{} consists of two major parts -- the central aggregator and the 5G mobile phone user or end-device; while the former has been implemented as a Python socket server, the latter has been implemented as a socket client. We have used the 5G datasets explained in Section~\ref{data} to represent ten end-users, of which two users use $50\%$ of the Lumos-5G dataset each, two users use $50\%$ of the Irish dataset each, two users use $50\%$ of the MN-Wild dataset each, and four users correspond to the Simulated-5G dataset. The prediction model has been implemented with the Keras module for importing the three model layers - LSTM, Dropout, and Dense, as shown in Table~\ref{tab:model_summary}. The initial training of the global model using the in-house 4G dataset and the subsequent local training, as well as retraining at the individual users using the aforementioned 5G datasets, are executed with a train-test split of $70\%$-$30\%$. While training the end-device's respective dataset, we have fixed the number of epochs as $25$. As Keras supports, the trained model is saved in a single HDF5 file containing the model's architecture, weight values, and compiled information. The size of this file for the central aggregator is $3.6$MB, while that at the end-devices is around $800$-$1300$KB.

\subsection{Baselines}
We used the following baselines.\\ 
\noindent\textbf{Raca2020}~\cite{Raca2020}: This work used an LSTM model to predict the 5G network throughput from various physical layer metrics, such as CQI, MCS, SINR, RSRP, etc., and user mobility info. As a baseline, keeping the common features intact, we implemented this model with two LSTM layers and each with a dropout of $0.2$, and a final Dense Layer.\\
\noindent\textbf{Minovski-RF}~\cite{Minovski2021}: This work uses different regression models, such as RF, Support Vector Regression, XGBoost, etc., to predict the 5G network throughput from network-level features. As a baseline, we have implemented an RF regression model with features, such as RSRP, SINR, CQI, etc., that the authors have used to develop their model. The heterogeneity in the 5G datasets might lead to model overfitting for the same set of hyperparameters, such as the number of estimators and max depth of the RF decision tree. Thus, we have kept these as default in the Python setup.\\
\noindent\textbf{Time Series Forecast (TS)}~\cite{raca2017back}: In~\cite{raca2017back}, the authors have explored different time series forecasting mechanisms, such as ARIMA and EWMA to predict the network throughput. The simple ARIMA version mentioned in~\cite{raca2017back} uses past samples of the throughput data. The exogenous features other than the target, i.e., throughput, are not considered for prediction here. We implement this ARIMA model as a baseline for our evaluation.

\begin{figure*}[!htb]
	\centering
	\subfigure[]{%
		\centering
		\label{fig:actual_vs_predicted_fedput}%
		\includegraphics[width=0.24\textwidth]{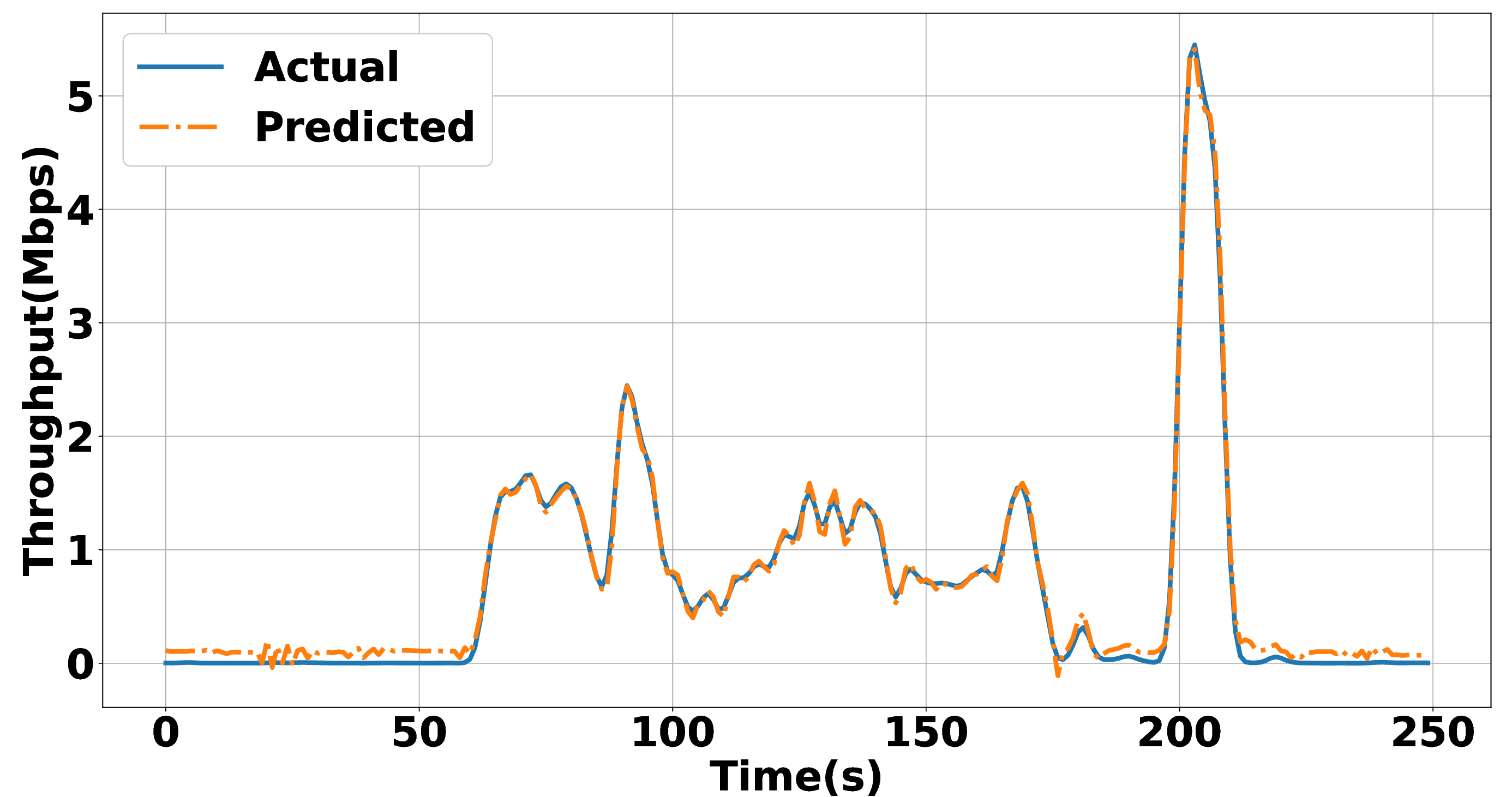}}\hfil
	\subfigure[]{%
		\centering
		\label{fig:actual_vs_predicted_lstm}%
		\includegraphics[width=0.24\textwidth]{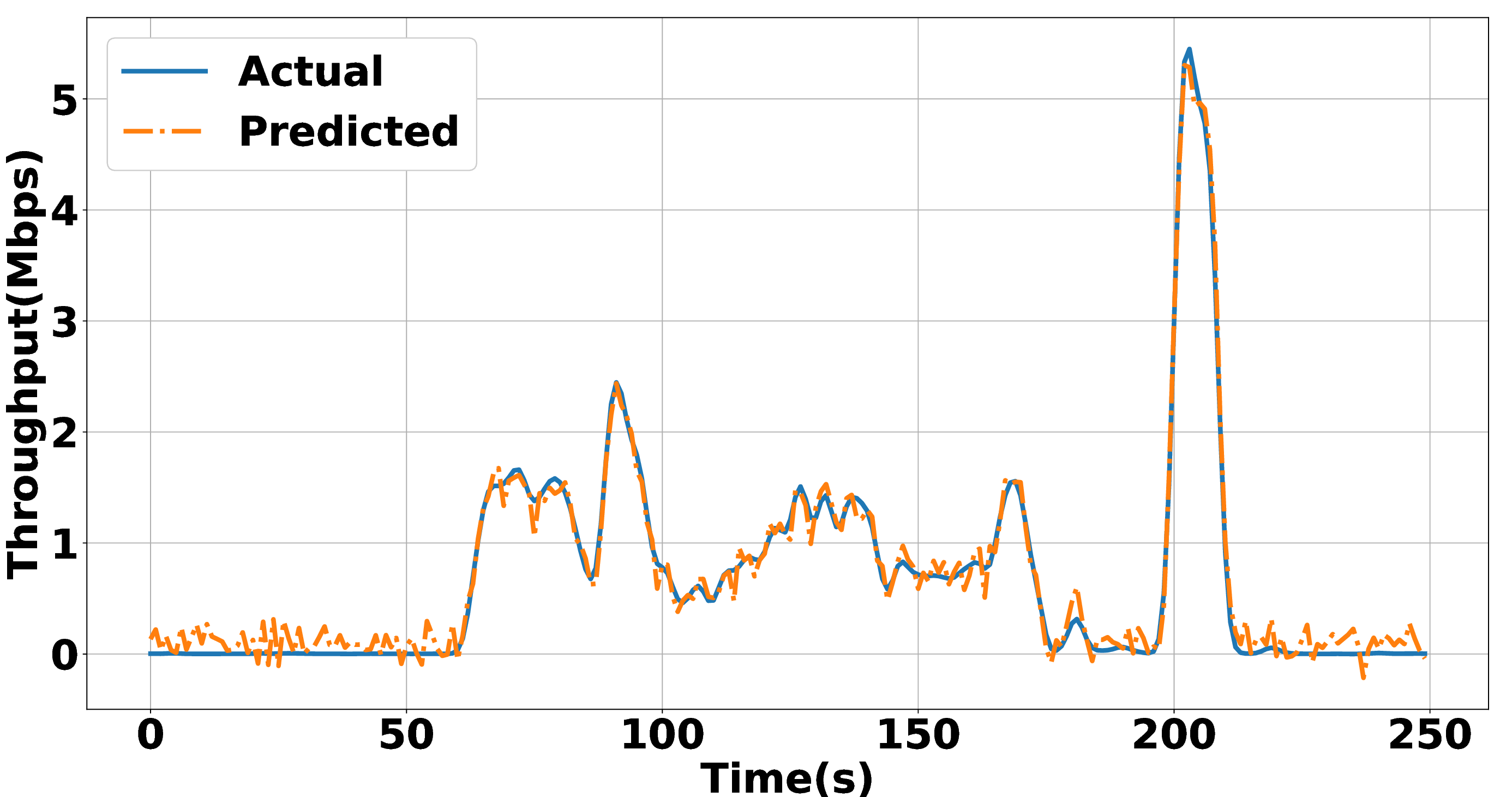}}\hfil
	\subfigure[]{
		\centering
		\includegraphics[width=0.24\textwidth]{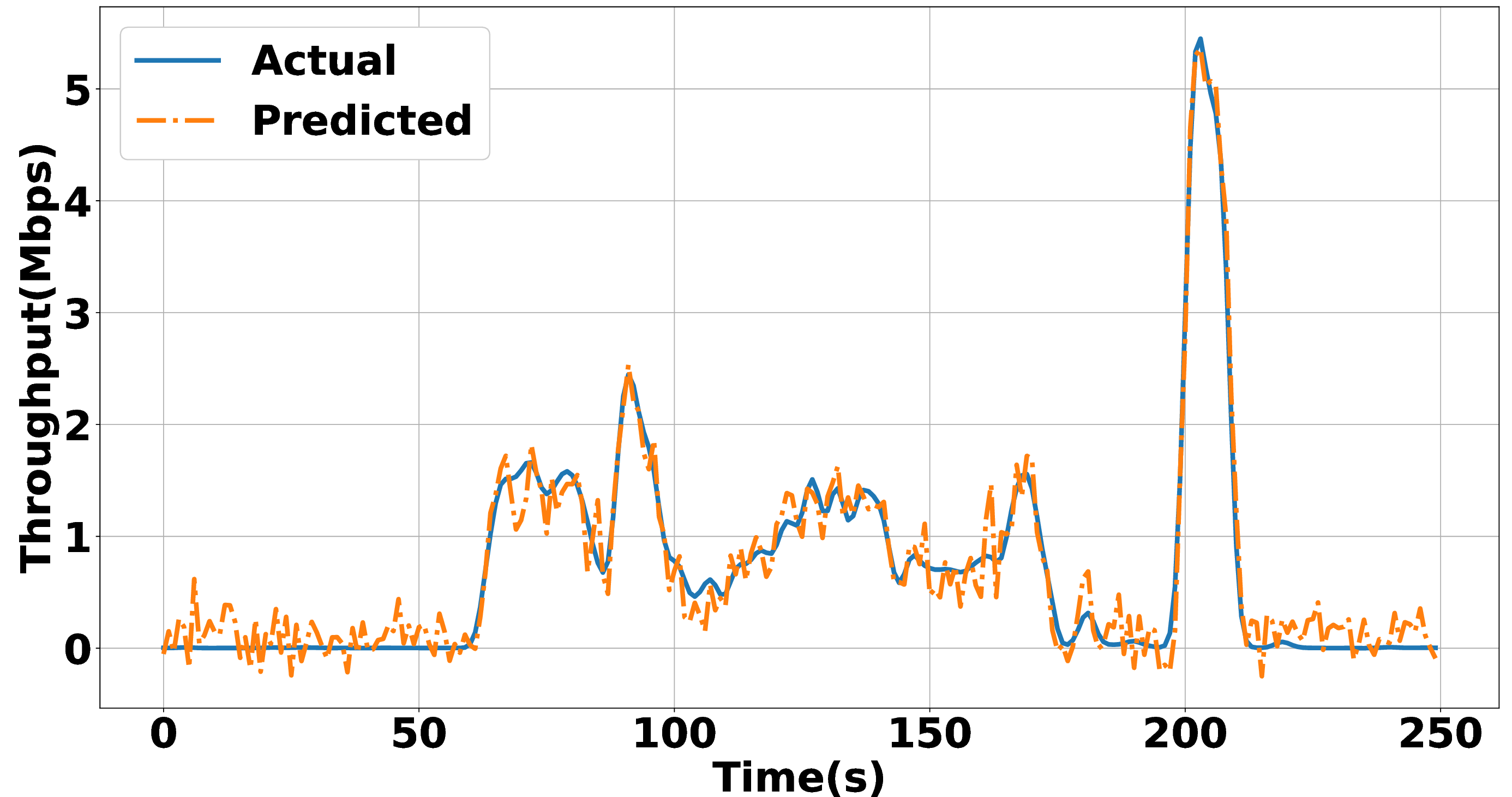}
		\label{fig:actual_vs_predicted_arima}}\hfil
	\subfigure[]{%
		\centering
		\label{fig:actual_vs_predicted_rf}%
		\includegraphics[width=0.24\textwidth]{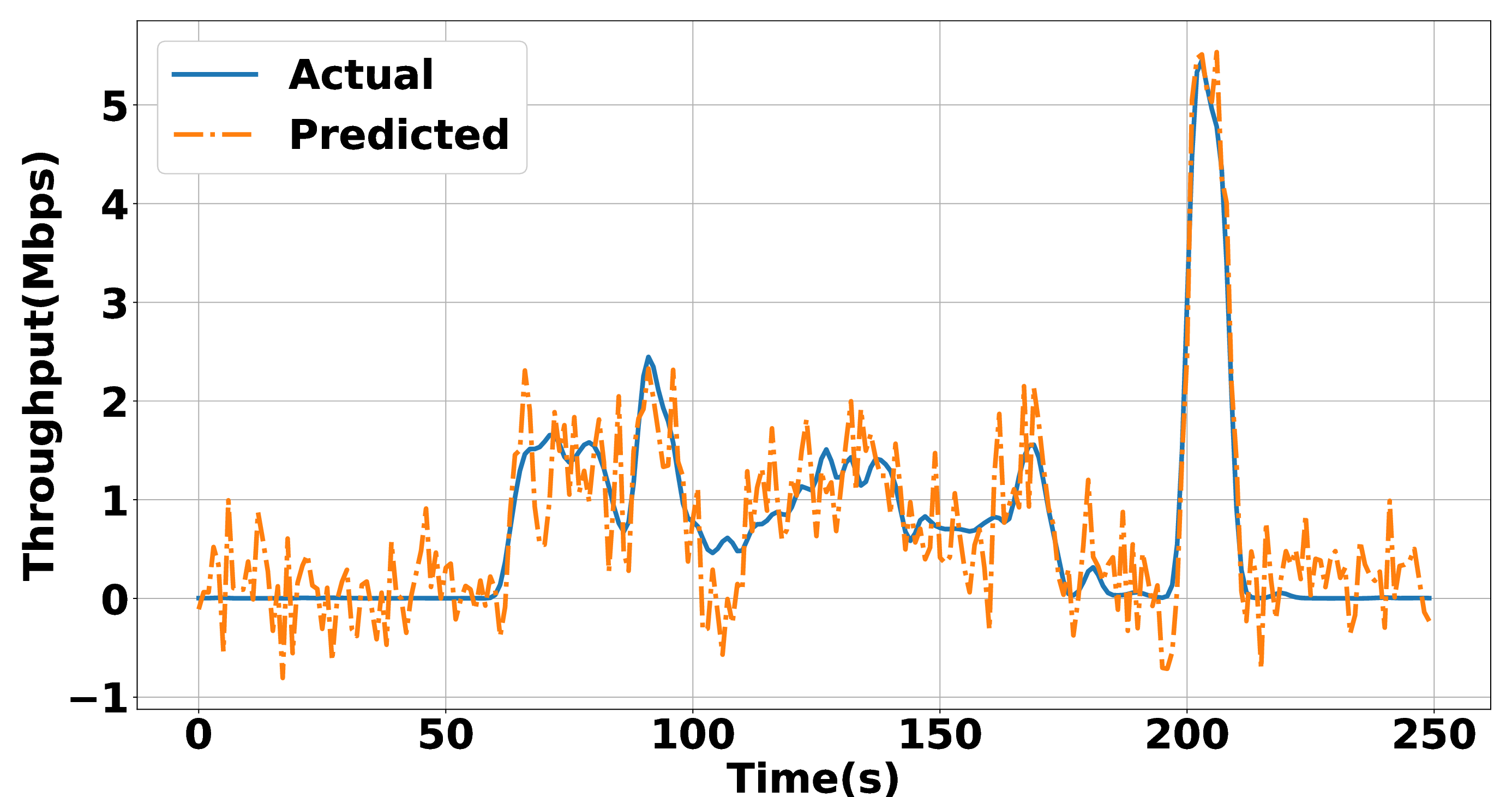}
	}
	\caption{Actual vs predicted throughput (a) using \ourmethod{}, (b) using \textbf{Raca2020} based model, (c) using \textbf{TS} based model, (d) using \textbf{Minovski-RF}}\label{fig:actual_vs_predicted}
\end{figure*}
\begin{figure*}[!htb]%
	\centering
	\subfigure[]{%
		\centering
		\label{fig:algo_comp_R_2}%
		\includegraphics[width=0.30\textwidth,keepaspectratio]{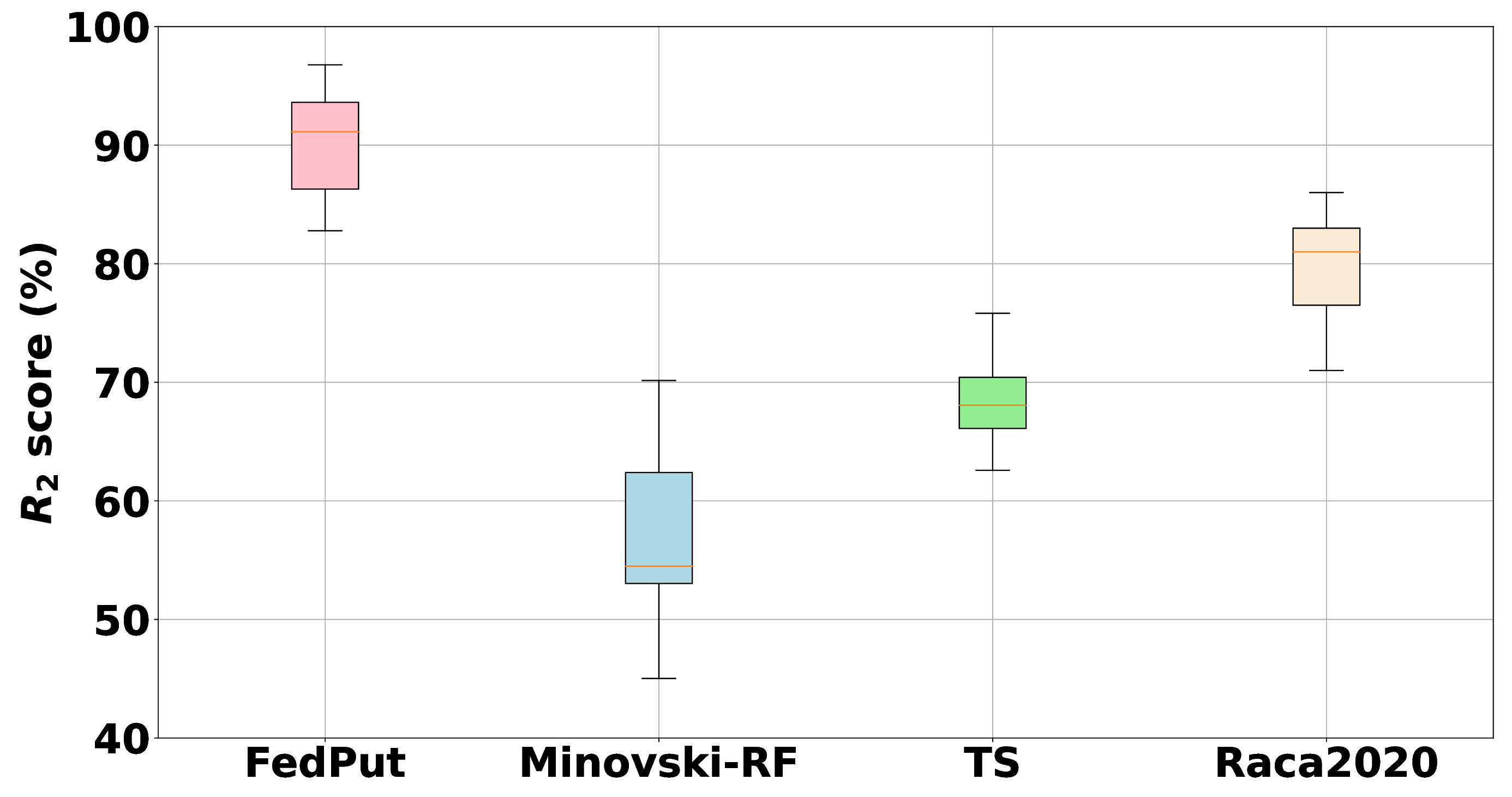}}%
	\subfigure[]{%
		\centering
		\label{fig:lumos_irish_pred}%
    \includegraphics[width=0.36\textwidth,keepaspectratio]{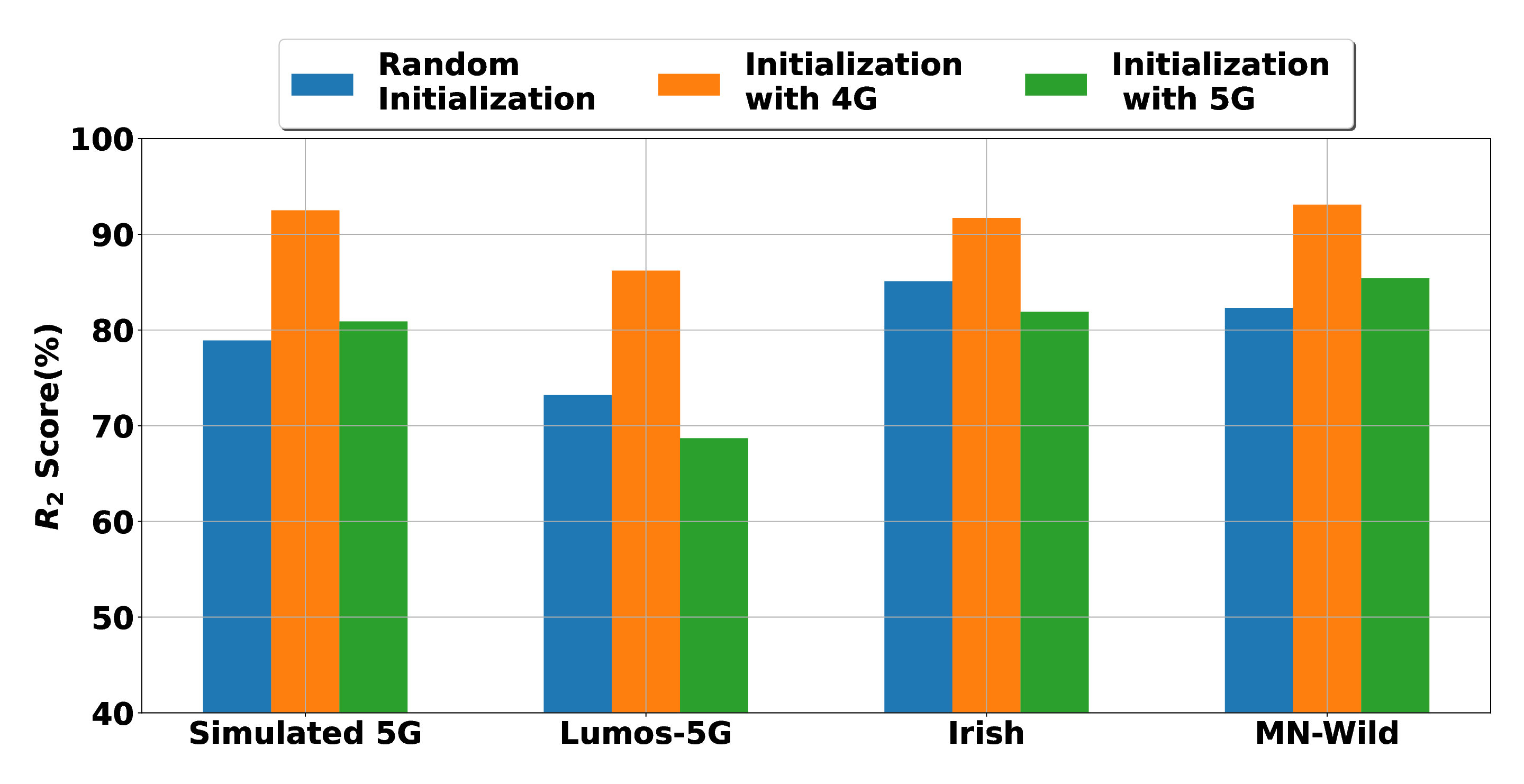}}
    \subfigure[]{
        \centering
        \label{fig:TT_R_2}
    \includegraphics[width=0.32\textwidth]{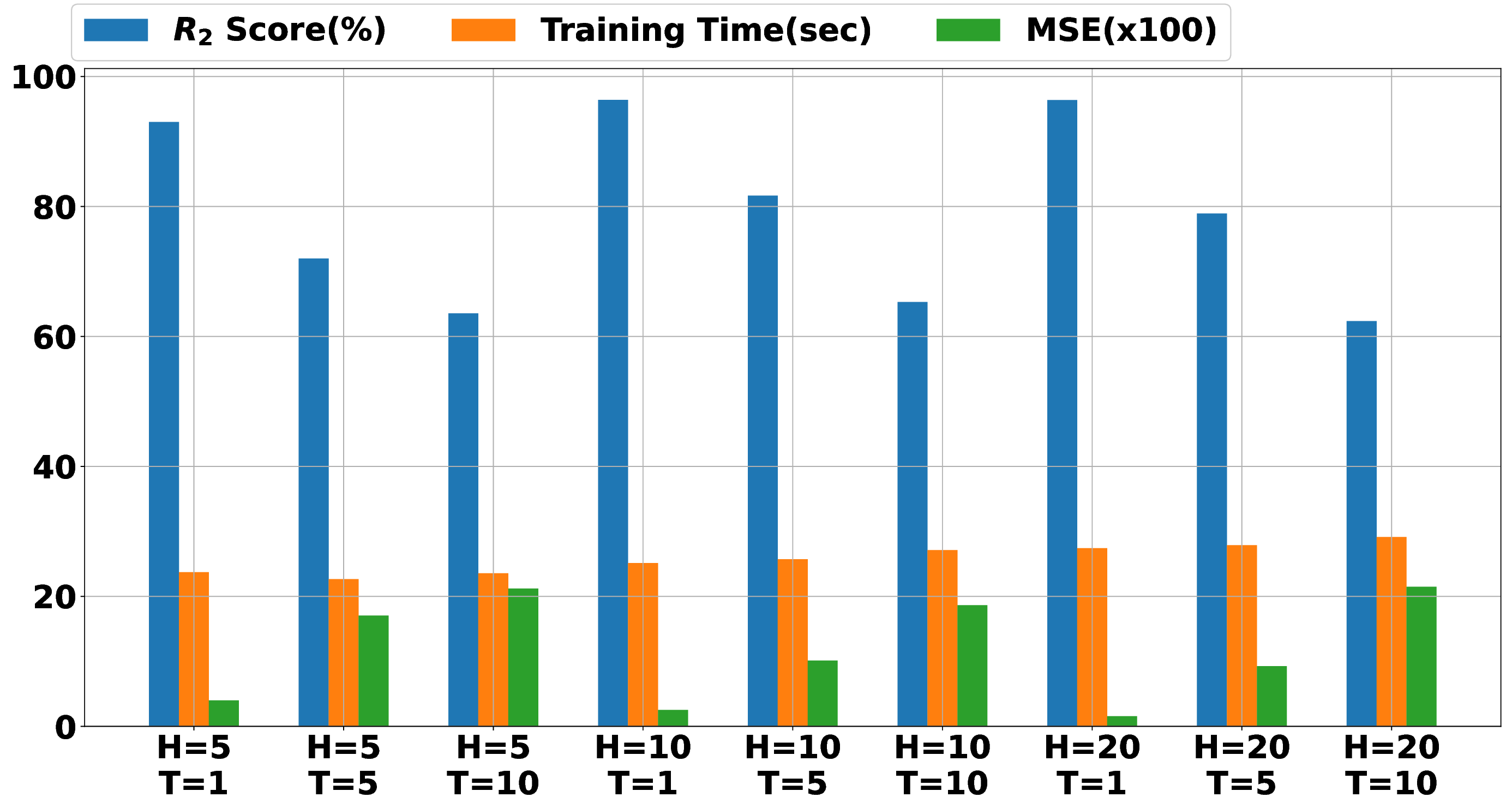}}
	\caption{$R_2$ score (a) for \ourmethod{}, Random Forest, ARIMA and LSTM (b) different initialization methods, and (c) $R_2$ score, MSE, and training time of \ourmethod{} algorithm for different history and prediction window size}
	\label{fig:r2score}
\end{figure*}

\subsection{Overall Performance}
We have evaluated and compared the performance of \ourmethod{} under two setups -- (a) with the simulated dataset and (b) with the publicly available 5G dataset. For \ourmethod{} implementation, we have used a history window size of $H=5$ seconds and a prediction window size of $W = 1$ second.

In \figurename~\ref{fig:actual_vs_predicted}, we visually compare the actual vs. predicted throughput for different algorithms with respect to time for the simulated 5G user. Compared to the other baselines, we observe that \ourmethod{} provides a closer match between the actual and the predicted throughput. To analyze further, \figurename~\ref{fig:algo_comp_R_2} shows the corresponding $R_2$ scores for different throughput predictors for the simulated 5G dataset. It is observed that while the average $R_2$ score of ARIMA and RF-based learning models is  69\% and 63.1\%, respectively, that of LSTM increases to 82\% and of \ourmethod{} increases to 91.4\%. It may, therefore, be inferred that our proposed \ourmethod{} based network throughput prediction algorithm achieves a reasonably high prediction accuracy. This is mainly because \ourmethod{} captures both the hardware as well network heterogeneity across all user devices and locations in a collaborative manner through the designed federated setup. The combined effect manifests in improved accuracy.

\subsection{Impact of Model Initialization}
In \figurename~\ref{fig:lumos_irish_pred}, we show the prediction accuracy for the four 5G datasets -- (i) Simulated-5G, (ii) Lumos-5G, (iii) Irish, (iv) MN-Wild. For each dataset, we have performed the prediction using three different initialization methods : (i) \textbf{Random model initialization} - here the central model is initialized with random weights, (ii) \textbf{4G bootstraping} - here the central model is initialized with weights obtained from the model trained with 4G data as discussed in Section~\ref{sec_methodology}, and (iii) \textbf{5G data trained model initialization} where the central model is initialized using the weights obtained from the four 5G datasets trained model. For this purpose, we have merged $70\%$ data from each 5G dataset and trained the model using our \ac{LSTM} framework. We evaluate the $R_2$ score of each 5G dataset for the rest of the $30\%$ data. Here we observe model initialization with the 4G dataset gives us better prediction accuracy. This is because the 4G dataset is collected across multiple locations using different mobile phone users and various network operators, making the model rich in information and more robust for model initialization. The central model initialized with the 5G dataset performs worse because - a) it relies on the simulated dataset and b) the real-world 5G datasets that have been collected from fewer geographical locations in comparison to the 4G dataset. Furthermore, the individual heterogeneity of the 5G datasets makes model prediction difficult.

\subsection{Impact of the Window Size}
First, we show the impact of the history window size ($H$) and the prediction window size ($T$) on the $R_2$ score and training time of the proposed \ourmethod{} algorithm in \figurename~\ref{fig:TT_R_2}. It is observed that keeping $H$ fixed, if we increase $T$, then the average $R_2$ score reduces. This is because the throughput is predicted for a long time into the future, which reduces its dependence on historical information. The training time is seen to increase slightly with the increase in the length of the history window or prediction window. This is due to the corresponding increase in the number of training data points. It is particularly noted that the proposed algorithm faithfully predicts the future throughput for a historical window of $H=5$ as well as $H=10$ seconds with an $R_2$ score of more than $90\%$. Further, an increase in $H$ does not improve the $R_2$ score any further. 

\subsection{Impact of Federated Iterations}
\begin{figure}
	\centering
		%
		\includegraphics[width=0.35\textwidth]{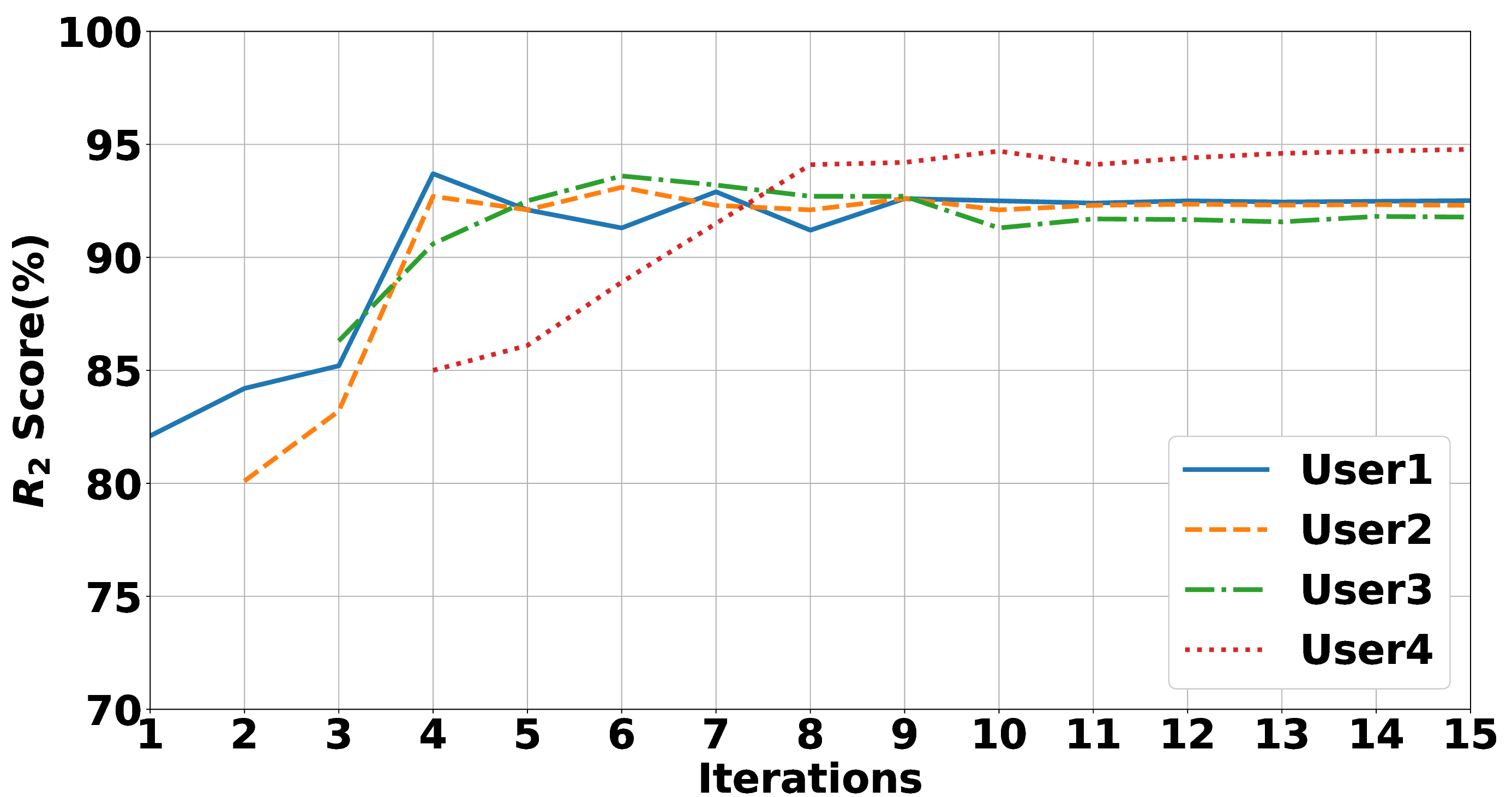}
	\caption{Convergence in the $R_2$ score while increasing the federated rounds and adding new users}
	\label{fig:r2ScoreIterWise}
\end{figure}

We have next simulated the performance of \ourmethod{} in a 5G mmWave setup as described in Section~\ref{public_data}, by increasing the number of federated iterations as well as the number of users. For this experimentation, we have used the four simulated 5G users. The result is shown in  \figurename~\ref{fig:r2ScoreIterWise}. In the first federated iteration, we used a single user. In the subsequent iterations, we have increased the number of users individually. It is observed from  \figurename~\ref{fig:r2ScoreIterWise} that with the gradual increase in the number of federated iterations, the $R_2$ score initially increases to a certain level and then saturates. 

\begin{figure*}%
	\centering
	\subfigure[]{%
		\centering
		\label{fig:qoe}%
		\includegraphics[width=0.33\textwidth]{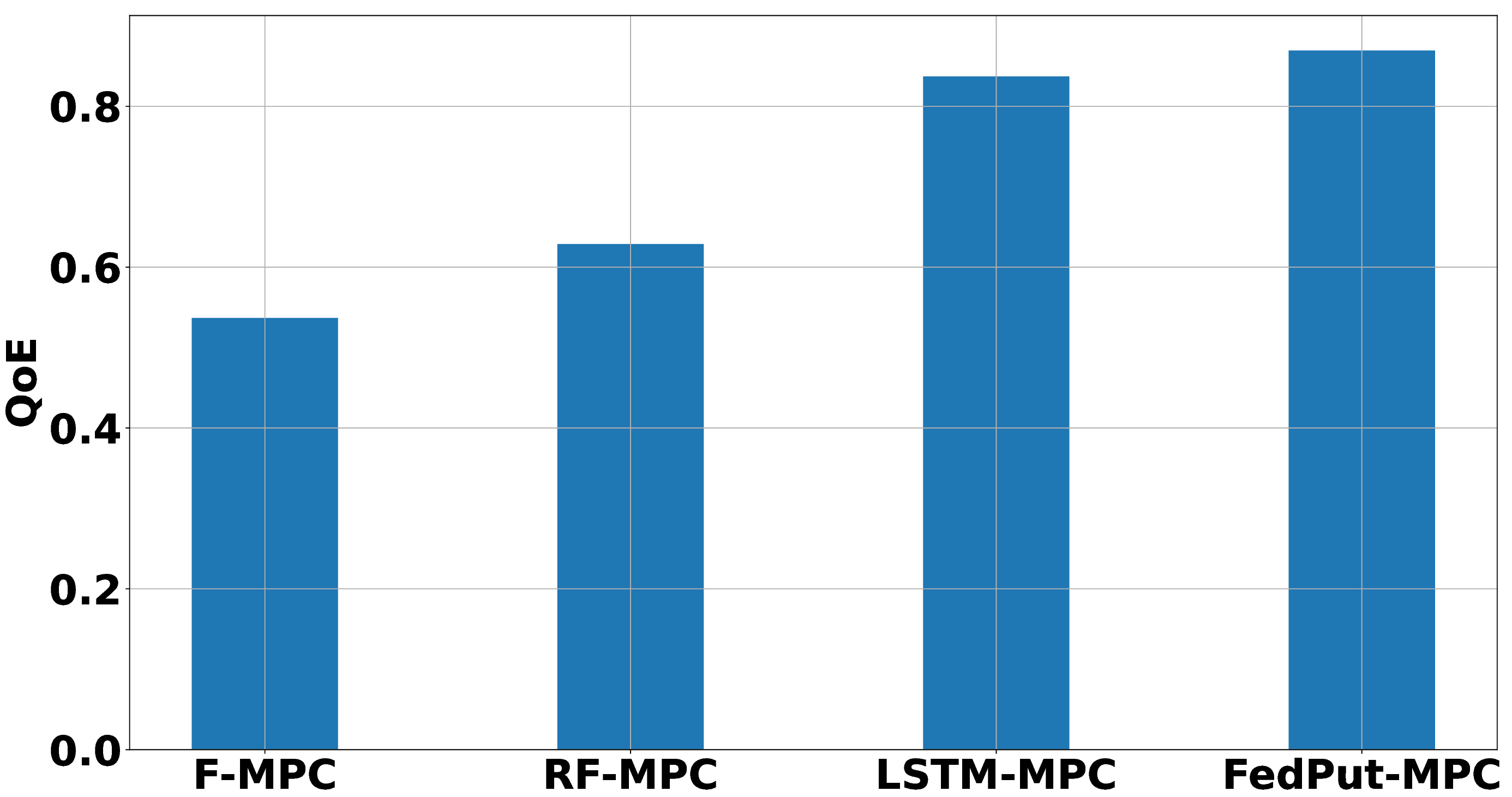}}\hfil
	\subfigure[]{%
		\centering
		\label{fig:bitrate}%
		\includegraphics[width=0.33\textwidth]{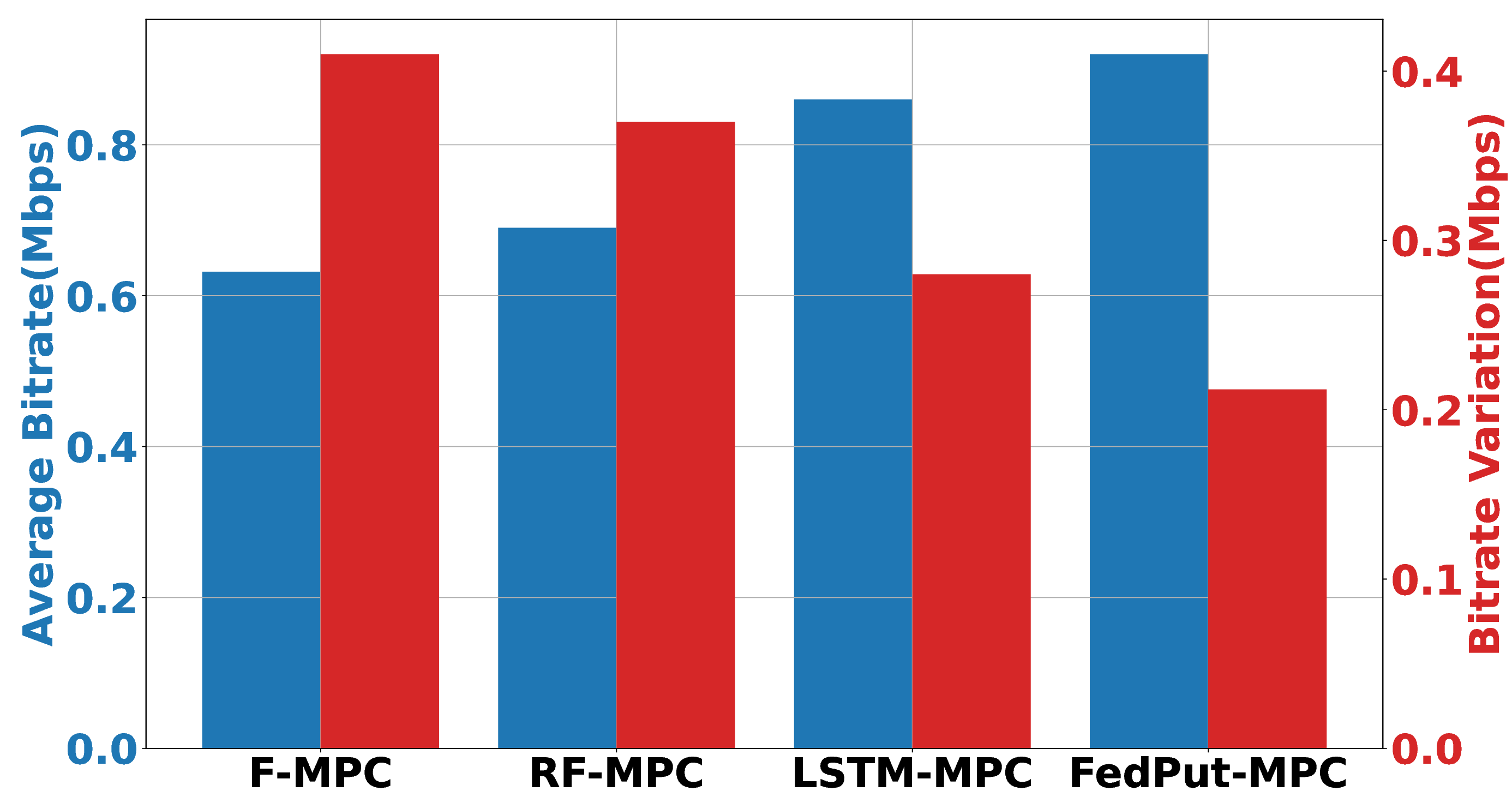}}
	\subfigure[]{%
		\centering
		\label{fig:rebuffering}%
		\includegraphics[width=0.33\textwidth]{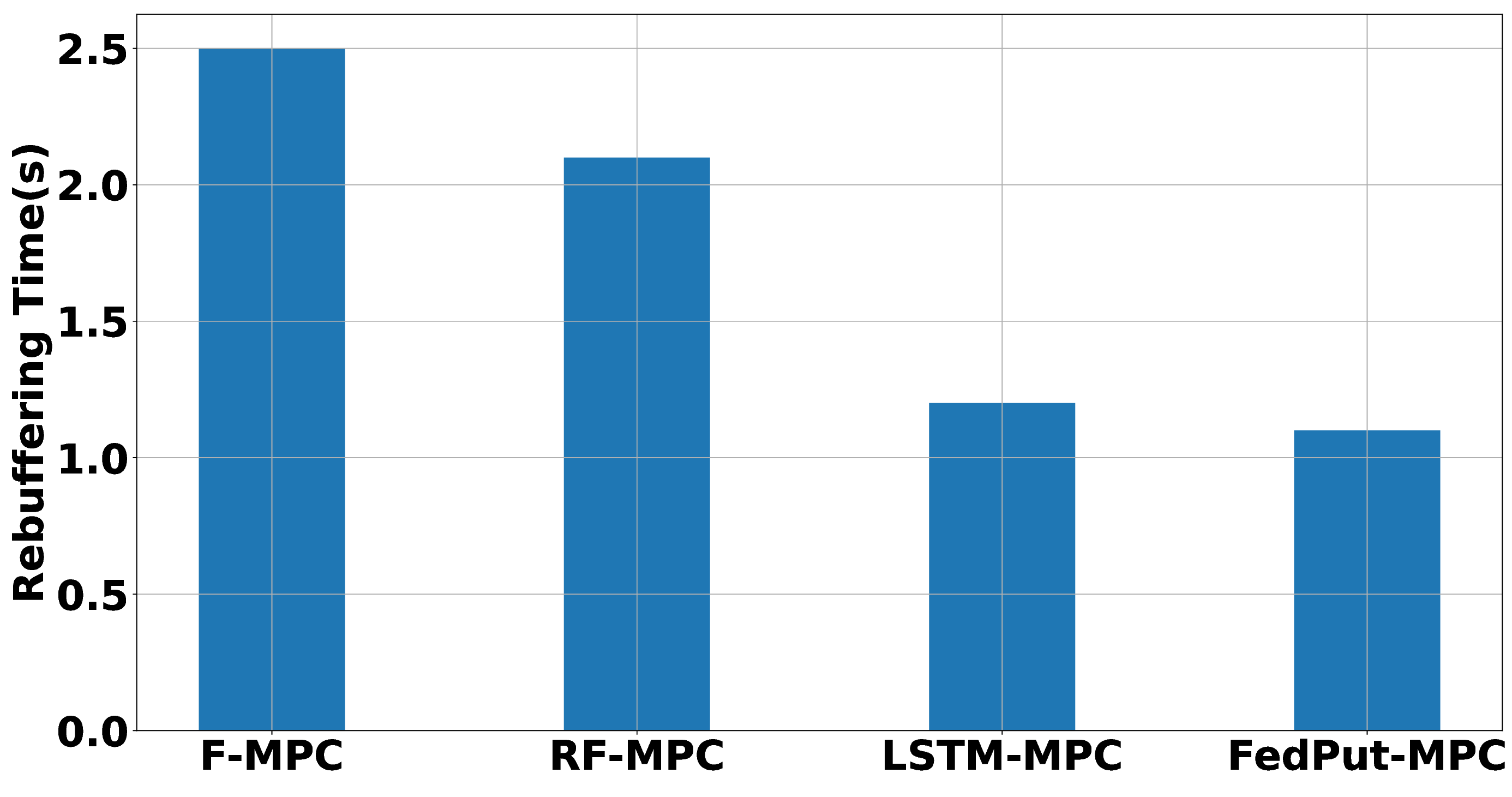}}
    \caption{(a) Video QoE under different ABR schemes, (b) average bitrate (Mbps) and bitrate variation (Mbps), (c) average rebuffering time per segment}
\end{figure*}
  \begin{figure*}
        \centering
	\subfigure[]{%
		\centering
		\label{fig:Qoecdf}%
		\includegraphics[width=0.24\textwidth]{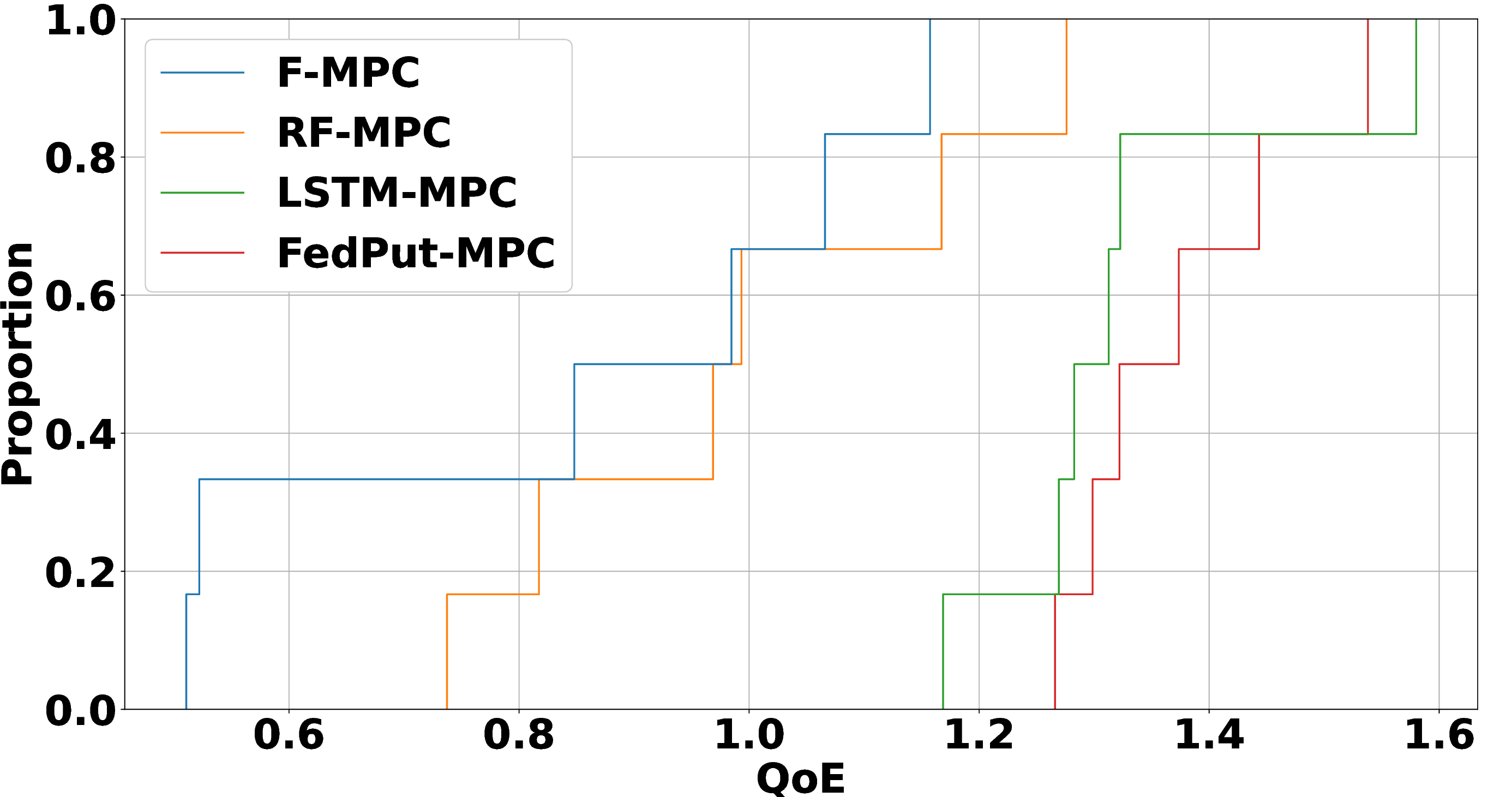}}\hfil
	\subfigure[]{%
		\centering
		\label{fig:Bitratecdf}%
		\includegraphics[width=0.24\textwidth]{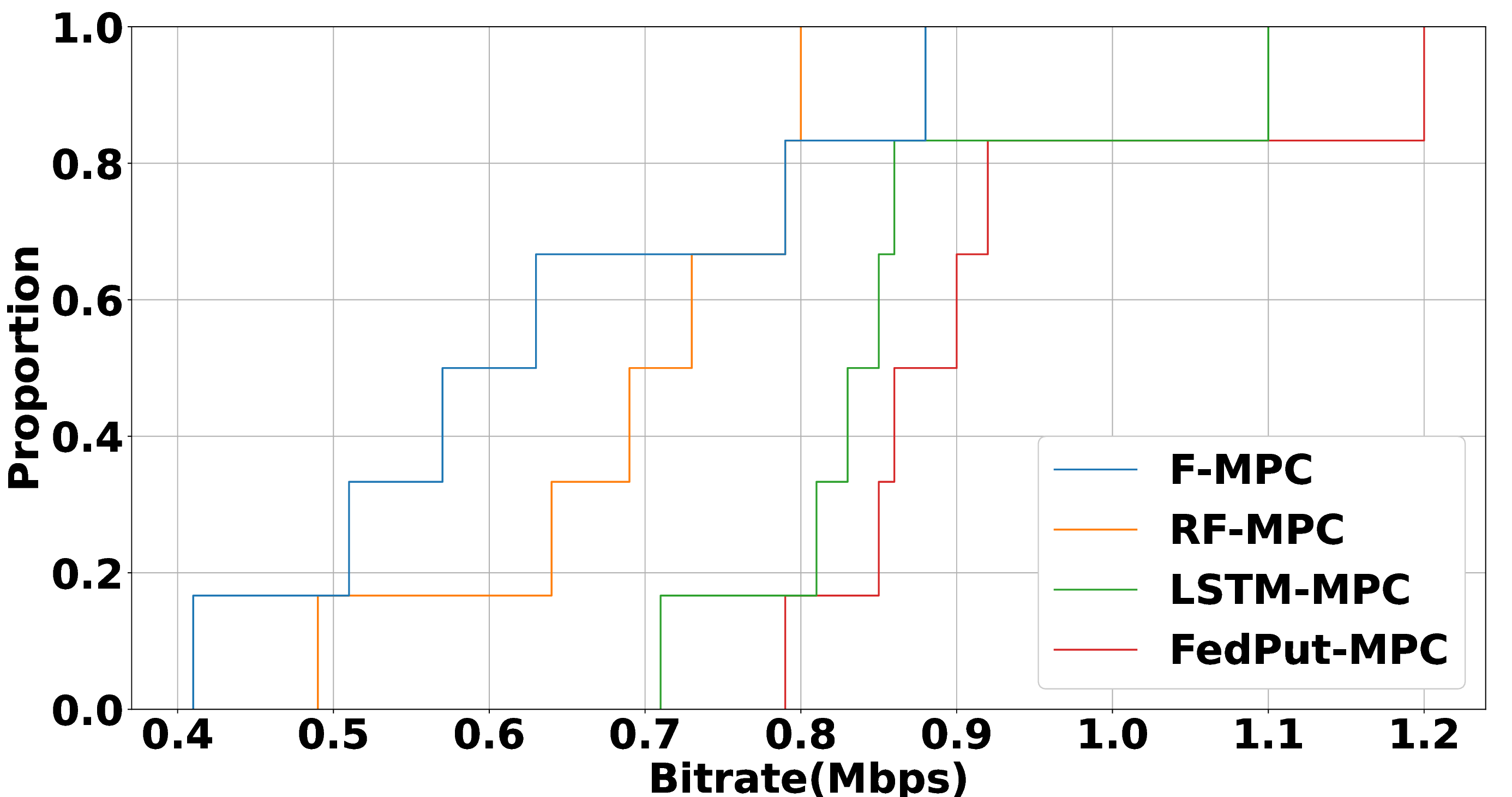}}\hfil
	\subfigure[]{%
		\centering
		\label{fig:BitrateVcdf}%
		\includegraphics[width=0.24\textwidth]{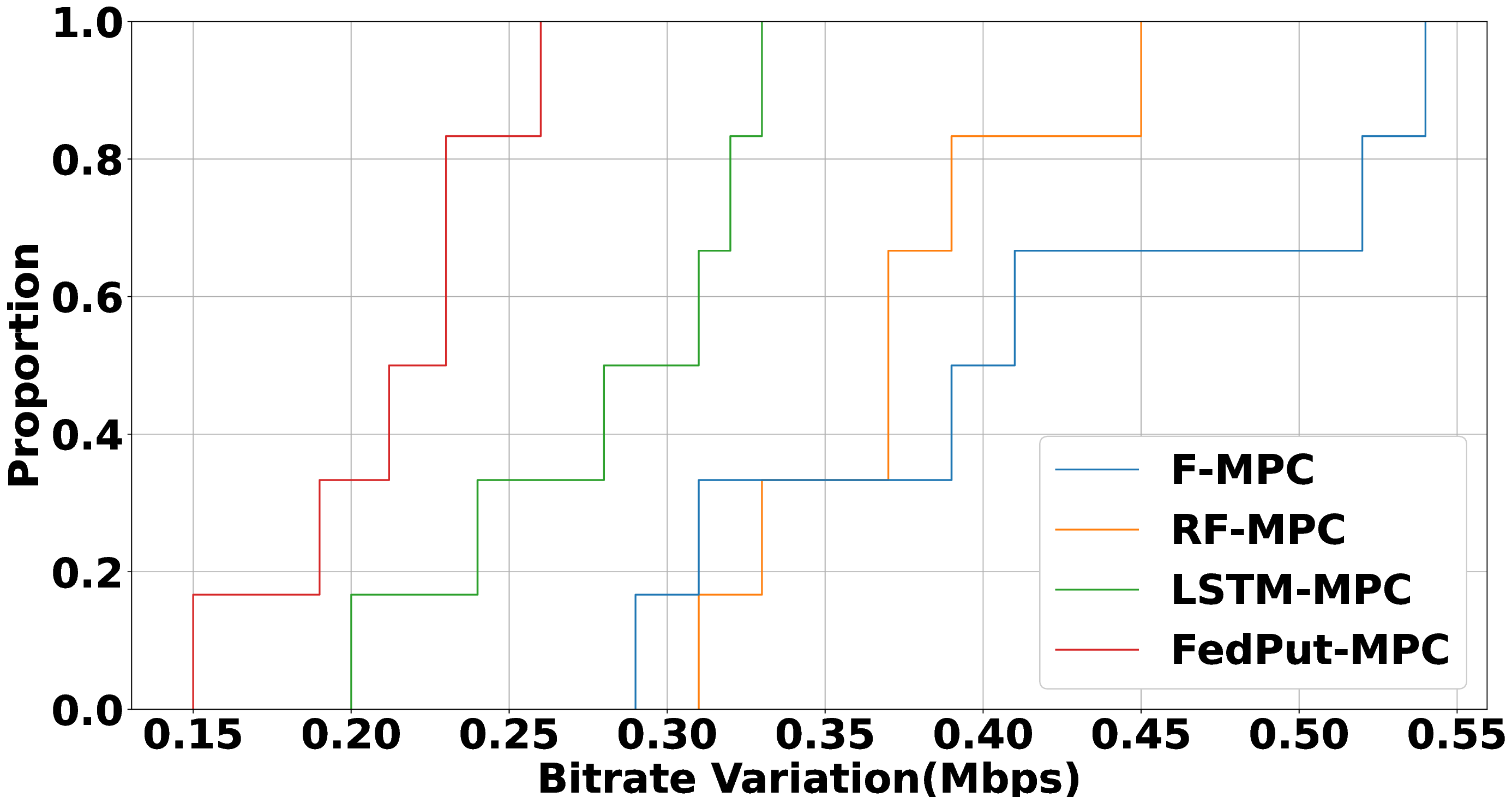}}\hfil
	\subfigure[]{%
		\centering
		\label{fig:RBFcdf}%
		\includegraphics[width=0.24\textwidth]{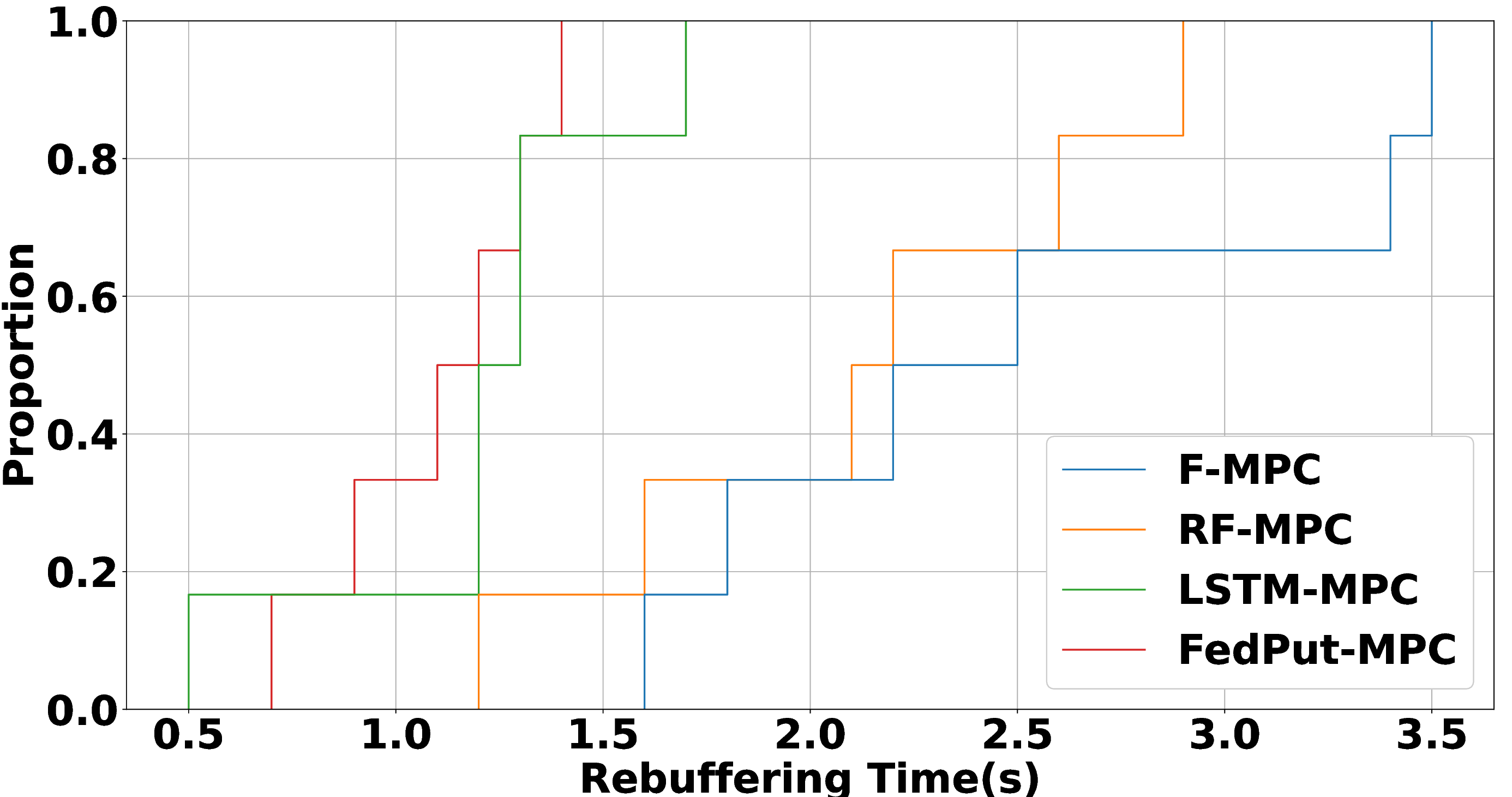}}\hfil
	\caption{Video QoE under different ABR schemes: (a) Empirical Cumulative Distribution Function (ECDF) of QoE under different ABR schemes,	(b) ECDF of avg. bitrate under different ABR schemes, (c) ECDF of bitrate variation under different ABR schemes, (d) ECDF of rebuffering time under different ABR schemes} 
	\label{fig:feature_importance2}
\end{figure*}


\subsection{PoC: Video Streaming Application}
A popular application that benefits from accurate throughput prediction is ABR media streaming from various video sources. We next evaluate how the QoE of ABR video streaming can be improved using \ourmethod{}.

\subsubsection{Integrating \ourmethod{} with ABR Algorithms}
A state-of-the-art ABR streaming algorithm that uses estimated network throughput for video quality prediction is fastMPC~\cite{Yin2015}. It uses the harmonic mean of the past throughput to predict the future throughput. Keeping the rest of the ABR algorithm of MPC intact, we have replaced the harmonic mean throughput predictor with different throughput prediction engines such as \ac{RF}, \ac{LSTM} as well as \ourmethod{} and evaluated the performance of these ABR streaming in terms of QoE. Thus, we have these four different ABR schemes -- (a) fastMPC, (b) RF-MPC, (c) LSTM-MPC, and (d) \ourmethod{}-MPC. The RF and LSTM-based throughput predictor uses 70\% of the simulated-5G dataset for training while \ourmethod{} is trained via \ac{FL}. The corresponding trained models are added as the MPC throughput predictor, and finally, we perform the video streaming simulation in our ns3-mmwave setup (Section~\ref{data_simulated}). 

\subsubsection{The Video Streaming Setup}
The ABR video streaming in the 5G network has been set up as a simulation in the $\mathtt{ns3-mmwave}$~\cite{Mezzavilla2018} module as outlined in Section~\ref{data_simulated}. A minimum of $2$ and a maximum of $10$ UEs under mobility have been deployed in the $1$ km $\times$ $1$ km ns3 simulation scenario, with their average speed varying between $5$m/s and $20$m/s. Correspondingly, there are four different scenarios, for each of which we have run ten simulation drops with different random seeds and a video length of $250$s. We have generated the simulation traces for a typical user only for each of the four scenarios, thereby generating four different UE datasets. The results in this section are the average over the $10$ drops for each scenario.

As mentioned in Section~\ref{data_simulated}, the throughput predictor is hosted as a Python socket server. It collects the location and network-related features from socket clients at the UE and predicts the network throughput for the future time window of $1$s. The predicted throughput value is passed to the socket client running at the ns3 UE, and based on the predicted throughput, the MPC ABR controller chooses the optimal chunk bitrate. The performance of the ABR algorithm has been evaluated using the generic \ac{QoE} metric~\cite[eqn. (5)]{Yin2015} with a video smoothness penalty of one and a video rebuffering penalty of $4.3$. The ABR controller can support six possible bitrate values, from $6.5$Mbps to $50$Mbps ($6.5$, $10$, $15$, $20$, $30$, $50$). 

\subsubsection{Results}
The QoE for each ABR scheme is shown in \figurename~\ref{fig:qoe}. The QoE metrics of the different algorithms are shown in \figurename~\ref{fig:qoe}. It may be observed that \ourmethod{}-MPC yields the highest average QoE compared to other variations. To further analyze the QoE performance, we have plotted each component of the QoE metric. \figurename \ref{fig:bitrate} shows the mean of the average bitrate and bitrate variation while \figurename \ref{fig:rebuffering} shows the average rebuffering time per video segment. From our observations, we find that the mean of the average bitrate (\figurename \ref{fig:bitrate}) of our proposed \ourmethod{} algorithm is comparable to LSTM-MPC; however, \ourmethod{} outperforms LSTM in terms of the rebuffering time per video segment of  (\figurename~\ref{fig:rebuffering}) as well as the bitrate variation. As rebuffering time and bitrate variation hurt the QoE estimation, thus, the QoE of our proposed algorithm is slightly better than LSTM-MPC. In \figurename~\ref{fig:Qoecdf}, we have shown the variation of the Empirical cumulative distribution function (ECDF) of QoE under the four ABR schemes. Our evaluation shows \ourmethod{}-MPC outperforms other schemes and achieves a higher QoE score. ECDF plots for individual components of QoE metric are also demonstrated in \figurename \ref{fig:Bitratecdf}, \ref{fig:BitrateVcdf}, \ref{fig:RBFcdf}. These explain clearly that the distribution of average bitrate is higher for \ourmethod{}-MPC predictor than the other predictors. The opposite distribution is observed for bitrate variation and rebuffering time. This is because \ourmethod{} provides a more robust throughput prediction that informs the video streaming application of network conditions. The average QoE of \ourmethod{}-MPC is $19.54\%$ higher than the SOTA F-MPC, compared to the closest competing LSTM-MPC predictor, \ourmethod{}-MPC gives $5.07\%$ higher average QoE.

\section{Conclusion}\label{sec_Conclusion}
In this work, we have proposed \ourmethod{}, a \ac{FL}  based throughput prediction algorithm for cellular networks. Unlike existing machine learning-based throughput prediction algorithms which are trained using centralized datasets, the proposed algorithm facilitates distributed training of a deep neural network-based model at end-devices. This addresses the issue that service providers may be reluctant to share their proprietary network information. Additionally, the use of \ac{FL} incorporates the user-specific variations of throughput into the prediction engine, which makes it suitable for a wide range of pervasive applications.  


\balance
\bibliographystyle{abbrv}
\bibliography{jsys.bib}
\balance
\end{document}